\documentclass[aps,prd,superscriptaddress,showpacs,floatfix,twocolumn,preprintnumbers,nofootinbib]{revtex4}

\usepackage{graphicx}
\usepackage{dcolumn}
\usepackage{amsmath,amssymb,bm,dsfont}
\usepackage{citesort}
\usepackage{slashed}
\usepackage{color}

\renewcommand{\Re}{\textrm{Re}}
\renewcommand{\Im}{\textrm{Im}}
\newcommand{\nn}{\nonumber}
\newcommand{\reff}[1]{(\ref{#1})}

\newcommand{\Lagr}{\mathcal{L}}
\newcommand{\MM}{\mathcal{M}^\mu}

\allowdisplaybreaks[1]
\begin{document}

\title{Pion photoproduction off the proton in a gauge-invariant chiral unitary framework}
\author{Maxim Mai}
\email{mai@hiskp.uni-bonn.de}
\affiliation{Universit\"at Bonn, Helmholtz-Institut f\"ur Strahlen- und Kernphysik (Theorie) and Bethe Center for Theoretical Physics, D-53115 Bonn, Germany}
\author{Peter C. Bruns}
\email{peter.bruns@physik.uni-regensburg.de}
\affiliation{Institut f\"ur Theoretische Physik, Universit\"at Regensburg, D-93040 Regensburg, Germany}
\author{Ulf-G. Mei{\ss}ner}
\email{meissner@hiskp.uni-bonn.de}
\affiliation{Universit\"at Bonn, Helmholtz-Institut f\"ur Strahlen- und Kernphysik (Theorie) and Bethe Center for Theoretical Physics, D-53115 Bonn, Germany}
\affiliation{Forschungszentrum J\"ulich, Institut f\"ur Kernphysik, Institute for Advanced Simulation, and J\"ulich Center for Hadron Physics, D-52425 J\"ulich, Germany}
\date{\today}

\begin{abstract}
We investigate pion photoproduction off the proton in a manifestly gauge-invariant chiral unitary extension of chiral perturbation theory.
In a first step, we consider meson-baryon scattering taking into account all next-to-leading order contact interactions. The
resulting low-energy constants are determined by a fit to s-wave pion-nucleon scattering and the low-energy data for
the reaction $\pi^- p\to \eta n$. To assess the theoretical uncertainty, we perform two different fit strategies. Having determined
the low-energy constants, we then analyse the data on the s-wave multipole  amplitudes  $E_{0+}$ of pion and eta photoproduction. These are parameter-free predictions, as the two new low-energy constants are
determined by the neutron and proton magnetic moments.  
\end{abstract}

\pacs{12.39.Fe, 13.75.Gx, 13.75.Jz, 14.20.Jn}

\maketitle
\section{Introduction and summary}

Pseudoscalar meson photoproduction off protons is one of the premier tools to unravel the spectrum and properties of
baryons made of the light up, down and strange quarks, as witnessed e.g. by the dedicated baryon resonance programs
at ELSA (Bonn) and CEBAF (Jefferson Laboratory).
Some of the low-lying resonances like the Roper $N^\star(1440)$,
the $S_{11}(1535)$ or the $\Lambda(1405)$ exhibit features that can not easily be reconciled with a simple constituent
quark model picture. Therefore, it was speculated since long that some of these peculiar states and their properties can
be explained if one assumes that they are generated through strong coupled-channel dynamics. Arguably the best
tool to address such a dynamical generation of resonances is unitarized chiral perturbation theory \cite{Kaiser:1995eg,Kaiser:1995cy,Bruns:2010sv,Oset:1997it,Oller:2000ma,Oller:2000fj,Lutz:2001yb,Ikeda:2012au,Mai:2012dt}.

Over the years, we have developed and a applied a gauge-invariant chiral unitary coupled-channel approach 
based on the leading order (LO) chiral effective Lagrangian of QCD to kaon \cite{Borasoy:2007ku} 
and eta photoproduction \cite{Ruic:2011wf}. To go beyond LO, one first has to refine the description of
meson-baryon scattering in this framework as the strong  hadronic final-state interactions are a crucial ingredient
in evaluating the complete photoproduction amplitudes. Therefore, in a recent work \cite{Bruns:2010sv} we have 
developed a framework to analyse meson-baryon scattering incorporating
next-to-leading order (NLO) contributions of the
chiral Lagrangian  \cite{Frink:2006hx}. Our scheme is based on the solution of the Bethe-Salpeter equation (BSE),
with a kernel derived from the contact terms of the NLO chiral Lagrangian. We have summed up the full infinite series
of Feynman diagrams generated by the BSE without resorting to any of the commonly made approximations as e.g. the
on-shell approximation. In this way we were able to reproduce successfully both s-waves of pion-nucleon scattering
($S_{11}$ and $S_{31}$) between the $\pi N$ and $\eta N$ thresholds. For higher energies we have observed that
only the resonance $S_{11}(1535)$ but not the $S_{31}(1620)$ could be described well. We have concluded that the
$S_{31}(1620)$ does not have a prominent dynamically generated content. As a matter of fact after fixing the
$S_{11}$ partial wave in the first energy region, i.e. $W_{{\rm cms}}<1.56\,\mathrm{GeV}$, the $S_{11}$ amplitude
for the higher energies came out in astonishing agreement with current partial wave analyses. The examination
of the complex energy plane showed that also the pole position of the second s-wave resonance, i.e.
the $S_{11}(1650)$, agrees rather well with those given by the particle data group \cite{Yao:2006px}. It is therefore
natural to extend this approach to s-wave photoproduction of pions, where a large data
basis already exists. This is done in this paper.

The main results of our investigation can be summarized as follows:
\begin{itemize}
\item[1)] As the first step, we have considered the meson-baryon scattering processes that are relevant for 
pion photoproduction. We consider six coupled channels, utilizing the chiral effective Lagrangian
at NLO. In the kernel of the underlying Bethe-Salpeter equation, we include all terms allowed by the
symmetries. We do not use the on-shell approximation that is common in most works on related subjects.
The crossed-channel contributions are not yet included  since the proper treatment of the corresponding diagrams 
in a framework comparable to the one presented here constitutes an unsolved problem.
\item[2)] To pin down the parameters of the approach (low-energy and subtraction constants), we
perform two fit strategies.   In strategy (I), we use as input the data on elastic $\pi N$ scattering in the $S_{11}$ and $S_{31}$
partial waves for energies in the range  $(m_p+M_\pi)<W_{{\rm cms}}<1.56$~GeV. In strategy (II), three subtraction constants are fixed and the
 $S_{11}$ partial wave is fitted up to $W_{{\rm cms}}<1.70$~GeV,  but the $S_{31}$ only in the near-threshold
 region, $W_{{\rm cms}}<1.20$~GeV.  The data on $\pi^- p\to \eta n$ from Ref.~\cite{Prakhov:2005qb} is included in both fit strategies.
\item[3)] In both fit strategies, the $S_{11}$ partial wave and the data on $\pi^- p\to \eta n$ are well described.
The $N^\star(1535)$ and the $N^\star(1650)$ are both dynamically generated, the precise pole positions depend
on the fit strategy, cf. Eqs.~(\ref{polesI},\ref{polesII}). 
We also give predictions for the scattering lengths $a_{\eta N\to \eta N}$ and $a_{\pi^-p\to \eta n}$.
\item[4)] Having scrutinized the hadronic sector, we have extended our approach to s-wave pion photo production. 
The only new parameters can be determined from the nucleon magnetic moments 
and thus parameter-free predictions emerge. We find a good description of the s-wave multipole $E_{0+}$ 
for pion photoproduction in the $S_{11}$-wave and also for $\eta$ photoproduction.
\end{itemize}

Having summarized the most important results of our study, it is important to briefly discuss possible
improvements of the method. First, the crossed channel dynamics has to be included properly.
This will allow e.g. to
get a better description of the near-threshold region in pion-nucleon scattering.
Unfortunately, the exact implementation of both crossing symmetry and unitary has not been possible so far in approaches based on Feynman diagrams,
 in contrast to other approaches as e.g. the one based on Roy-Steiner equations \cite{Ditsche:2012fv} where these constraints are met by construction.
For an attempt to approximately restore crossing symmetry in an ansatz comparable to the one employed here, see e.g. \cite{Gasparyan:2010xz}.
Further, in some
channels, explicit resonance degrees of freedom will have to be incorporated as not all resonances
are generated dynamically.  For a method to do that, see e.g. Ref.~\cite{Meissner:1999vr}.
Finally, a larger data base including also kaon-nucleon scattering and kaon photoproduction should be
considered simultaneously with the processes studied here. Work along these lines is in progress.

 The paper is organized as follows: In Sec.~\ref{sec:hadron}, the underlying approach to analyse meson-baryon
 scattering is described. Two fit strategies to pin down the occurring parameters are presented and fits to data
 and predictions for some scattering lengths and the fundamental $\eta N$ scattering amplitude are given. The
 approach is extended to photoproduction in Sec.~\ref{sec:Photo}, where parameter-free predictions for pion
 and $\eta$ production off protons are given. Many technicalities are relegated to the appendices.

\section{Hadronic scattering}
\label{sec:hadron}

In the present section we wish to describe our framework for meson-baryon scattering as it will serve us as the most important ingredient in our analysis of the photoproduction processes. We wish also to give a more systematic error analysis of our results as it was done in \cite{Bruns:2010sv}. Moreover from this first study we know that our coupled channel approach is applicable to relatively high energies, thus we will extend our analysis also to the process $\pi N\rightarrow \eta N$.

\subsection{Formalism}

Chiral Perturbation Theory (ChPT) is a powerful tool to analyse pion-nucleon
scattering and pion photoproduction at low energies in a systematic manner. In
its original formulation one starts from the chiral Lagrangian. The reaction
amplitudes are then given by a set of diagrams including contact interactions
as well as loop diagrams. Working order by order in a low-energy expansion,
every Feynman graph can be classified by the order to which it contributes in
this expansion. For a specific chiral order the number of such diagrams is
finite. Thus technical effort for a calculation of the scattering amplitudes
can be large but remains finite. Up to now the scattering amplitudes for
meson-baryon scattering have been calculated in the three-flavour formulation
of ChPT up to the third chiral order in
Refs.~\cite{Mai:2009ce,Liu:2006xja}. However, the region where the result of
this strictly perturbative approach yields a decent description of the process
is restricted to the subthreshold region, and it certainly fails in the
vicinity of resonances (as nicely discussed e.g. in Ref.~\cite{Kaiser:2001hr}).

In the following we present a framework to analyse meson-baryon scattering for energies up to the resonance region. It is inspired by Chiral Perturbation Theory, which gives the driving terms of the meson-baryon interaction, and it relies on the coupled channel Bethe-Salpeter equation \cite{Salpeter:1951sz}. The latter implements the requirement of two-body unitarity exactly and in principle allows to generate resonances dynamically. It already improved our understanding of the purely mesonic and meson-baryon sector in recent years, see e.g. \cite{Kaiser:1995eg,Kaiser:1995cy,Oset:1997it,Oller:2000ma,Oller:2000fj,Lutz:2001yb,Ikeda:2012au,Mai:2012dt,Bruns:2010sv}.

\begin{figure}[t]
\begin{center}
\includegraphics[width=1.0\linewidth]{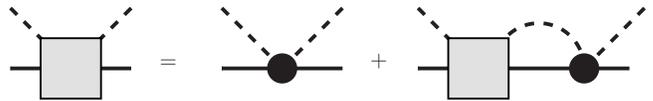}
\end{center}
\caption{Symbolical representation of the Bethe-Salpeter equation. Here the
  circle and the square  represent the potential $V$ and the scattering 
  amplitude $T$, respectively.}\label{pic:BSE}
\end{figure}

We denote the in- and out-going meson momenta by $q_1$ and $q_2$, respectively. Moreover the overall four-momentum is given by $p=q_1+p_1=q_2+p_2$, where $p_1$ and $p_2$ are the momenta of in- and out-going baryon, respectively. For the meson-baryon scattering  amplitude $T(q_2, q_1; p)$ and potential $V(q_2, q_1; p)$ the integral equation to solve reads in $d$ dimensions
\begin{align}\label{eqn:BSE}
      T({q}_2, &{q}_1; p)= V(q_2, q_1; p) \\
      &+ i\int\frac{d^d l}{(2\pi)^d}V({q}_2, {l}; p) S(\slashed{p}-\slashed{l})\Delta(l)T({l}, {q}_1; p) ,\nn
\end{align}
where $S$ and $\Delta$ represent the baryon (of mass $m$) and the meson (of mass $M$) propagator, respectively, and are given by $iS(\slashed{p}) ={i}/({\slashed{p}-m+i\epsilon})$ and $i\Delta(k) ={i}/({k^2-M^2+i\epsilon})$. Let us note that all objects in the above equation which are denoted by capital roman letters, are matrices in the channel space. Restricting the meson-baryon states to have the quantum numbers of the proton, the channel space reduces to the 6-dimensional space spanned by the following channels $\{\pi^0 p, \pi^+ n, \eta p, K^+ \Lambda, K^+\Sigma^0, K^0 \Sigma^+\}$. As a matter of fact, the BSE can be visualized as presented in Fig.\ref{pic:BSE}. Note that the loop integration in Eq.~\reff{eqn:BSE} is performed in $d$ dimensions, without restricting the loop momenta to be on the mass shell. Such an approximation would certainly reduce the technical effort to solve the BSE, however it spoils the direct correspondence of the solution of the BSE Eq.~\reff{eqn:BSE} to the series of Feynman graphs, which we evaluate as it stands. Thus, every term in our solution of the BSE is related, in a one-to-one correspondence, to a properly evaluated Feynman diagram. It comes as an advantage of this prescription that the implementation of gauge invariance in a combined analysis of hadronic scattering and  meson photoproduction is straightforward and very natural: It follows essentially the guidelines from quantum field theory textbooks, see e.g sec.~(7.4) of \cite{Peskin}, and will be explained in sec.~\ref{sec:Photo}. Moreover, it is also straightforward to compare our amplitude to the perturbative expansion at any fixed chiral order. We complete our discussion of the off-shell-dependence of the effective vertices with the remark that the use of the on-shell-approximation is not more ``physical'' than taking the off-shell-dependence into account, though reducing the effort of the calculation significantly down to the evaluation of a geometric series. Simply iterating a fixed on-shell kernel in such a geometric series can even lead to significant deviations from the results of Feynman graphs when iterating Born-terms, as is exemplified by an analysis of box graphs in sec.~(5.2) of \cite{RobinsDiss}. 
 In our case, off-shell behavior of the potential reflects itself in tadpole - integral terms in the full scattering amplitude. 
 These terms might in general depend on the chosen parameterization of fields
 as long as only a subset of Feynman graphs is summed up. Setting them to zero
 as done using the on-shell approximation is just one possible ``choice of
 gauge'' (in the space of field parameterizations) in a non-invariant result,
 which however is not in line with the proper evaluation of loop diagrams we
 aim at here. As the analytic energy-dependence of the tadpole-terms is
 trivial, it should be possible to compensate for this non-invariance effect
 by an adjustment of coupling constants in the kernel. As a byproduct
 of our procedure for finding fits\footnote{See the discussion in sec.~3 of
   \cite{Mai:2012dt}}, we have checked numerically that this is indeed the case.
Thus, while the exact numerical values of the coupling constants should be taken with a grain of salt (they should be considered as model parameters in our approach) the overall properties of the amplitude are solely based on the unambiguous analytic properties of the selected infinite subset of loop graphs.

\subsection{Interaction kernel}

The only missing part of equation \reff{eqn:BSE} remains the driving term of the meson-baryon interaction, i.e. the potential $V(q_2, q_1; p)$. It is well known that already the leading order chiral potential, the so-called Weinberg-Tomozawa term, captures the prominent part of the meson-baryon dynamics at the $\pi N$ threshold. It is derived from the covariant derivative of the leading order chiral Lagrangian, which reads
\begin{align}\label{eqn:LAGR0}
  \Lagr^{(1)}_{\phi B}&=\langle \bar{B} (i\gamma_\mu D^\mu-m_0)B\rangle+\frac{D/F}{2}\langle \bar{B}\gamma_\mu \gamma_5[u^\mu,B]_\pm \rangle ,
\end{align}
where $\langle\ldots\rangle$ denotes the trace in flavor space, $[D_\mu, B]
:=\partial_\mu B +\frac{1}{2}[[u^\dagger,\partial_\mu u],B]$, $m_0$ is the
common baryon octet mass in the chiral limit, while $D$ and $F$ are the lowest-order axial coupling constants. The relevant degrees of freedom of ChPT are the Goldstone bosons, which are collected in the traceless meson matrix $U\in SU(3)$,
\begin{equation}
  U =\exp\Bigl(i\frac{\phi}{F_0}\Bigr),  \phi=\sqrt{2}  \begin{pmatrix}  \frac{\pi^0}{\sqrt{2}}+\frac{\eta}{\sqrt{6}} \!\!&\!\! \pi^+ \!\!&\!\! K^+		\\
  \pi^- \!\!&\!\! -\frac{\pi^0}{\sqrt{2}}+\frac{\eta}{\sqrt{6}} \!\!&\!\! K^0											\\
  K^- \!\!&\!\! \bar{K}^0 \!\!&\!\! -\frac{2}{\sqrt{6}}\eta  \end{pmatrix},
\end{equation}
where $F_0$ is the meson decay constant in the chiral limit. The baryonic fields are collected in a traceless matrix
\begin{eqnarray}
  B=\begin{pmatrix}  \frac{\Sigma^0}{\sqrt{2}}+\frac{\Lambda}{\sqrt{6}} & \Sigma^+ & p		\\
  \Sigma^- & -\frac{\Sigma^0}{\sqrt{2}}+\frac{\Lambda}{\sqrt{6}}& n				\\
  \Xi^-& {\Xi}^0 & -\frac{2}{\sqrt{6}}\Lambda  \end{pmatrix}.
\end{eqnarray}
In view of the application to meson-baryon scattering we set external currents to zero except for the scalar one, which is set equal to the quark mass matrix, $s=\mathcal{M}:=\textrm{diag}(m_u, m_d, m_s)$.  Note that this situation will change as soon as we will deal with photoproduction in the next section. We furthermore use
\begin{align}
  u^2:=U  ,\quad  u^\mu:=iu^{\dagger}\partial^\mu u - iu\partial^\mu u^{\dagger} ,\nn		\\
  \quad\chi_\pm:=u^{\dagger}\chi u^{\dagger}\pm u\chi^{\dagger}u ,\quad  \chi:=2B_0 s,
\end{align}
where the constant $B_0$ is related to the quark condensate in the chiral limit. Separating the channel space from the momentum space structures, the Weinberg-Tomozawa potential reads
\begin{align}\label{VWT}
 V_{WT}(q_2, q_1; p)=A_{WT}(\slashed{q_1}+\slashed{q_2}),
\end{align}
where $A_{WT}$ denotes a matrix in channel space as defined in App. \ref{app:coupling}. The axial coupling in the Lagrangian \reff{eqn:LAGR0} induces an additional contribution to the chiral potential formally at the leading chiral order. It arises via the one-baryon exchange graphs in the $u$ and $s$-channel, the so-called Born graphs. 
It is not known how to fully include such terms in the scattering kernel of the BSE when attempting an analytic solution in $d$ dimensions, corresponding to Feynman graphs. The difficulties one encounters in this case are described in \cite{Bruns:2010sv}. On the other hand, it is plausible that the s-wave contributions stemming from the Born terms can be well approximated by a series of contact terms not too far above the two-particle thresholds. We shall return to this open problem in a separate publication \cite{MBM_2027}. In the present work, we will restrict the driving term of our coupled channel analysis to a set of contact terms.
We go beyond the leading order and complete this set by the meson-baryon vertices from the second order chiral Lagrangian from \cite{Krause:1990xc,Oller:2006yh,Frink:2004ic}. It reads
\begin{align}\label{eqn:LAGR1}
 &\Lagr^{(2)}_{\phi B}=~b_0 \langle\overline B B\rangle \langle\chi_+\rangle + b_{D/F} \langle\overline B \big[\chi_+,B\big]_\pm\rangle+b_4 \langle\overline B  B\rangle \langle u_\mu u^\mu\rangle 			\nn\\
 &+b_{1/2} \langle\overline B  \Big[u_\mu,\big[u^\mu,B\big]_\mp\Big]\rangle + b_3 \langle\overline B \Big\{ u_\mu,\big\{ u^\mu,B\big\}\Big\}\rangle									\nn\\
 &+ib_{5/6} \langle\overline B \sigma^{\mu\nu} \Big[\big[u_\mu,u_\nu\big], B\Big]_\mp\rangle + ib_7\langle\overline B \sigma^{\mu\nu} u_\mu\rangle  \langle u_\nu B\rangle						\nn\\
 &+ \frac{i\,b_{8/9}}{2m_0}\Big( \langle\overline B \gamma^\mu\Big[u_\mu,\big[u_\nu,\big[D^\nu, B\big]\big]_\mp\Big]\rangle												\nn\\
 &\hspace{+3.5cm}+\langle\overline B \gamma^\mu\Big[D_\nu,\big[u^\nu, \big[u_\mu,B\big]\big]_\mp\Big]\rangle\Big) 													\nn\\
 &+\frac{i\,b_{10}}{2m_0}\Big( \langle\overline B \gamma^\mu\Big\{ u_\mu,\big\{ u_\nu,\big[D^\nu,B\big]\big\}\Big\}\rangle												\nn\\
 &\hspace{+3.5cm}+\langle\overline B\gamma^\mu\Big[D_\nu,\big\{ u^\nu, \big\{ u_\mu,B\big\}\big\}\Big]\rangle\Big)													\nn\\
 &+\frac{i\,b_{11}}{2m_0}\Big( 2\langle\overline B \gamma^\mu \big[D_\nu,B\big]\rangle \langle u_\mu u^\nu\rangle													\nn\\
 &\hspace{+2.2cm}+\langle\overline B \gamma^\mu B\rangle \langle\big[D_\nu,u_\mu\big]u^\nu + u_\mu \big[D_\nu,u^\nu\big]\rangle   \Big)											\nn\\
 &+\Big( b_{12}\langle\overline B \sigma_{\mu\nu} [f_+^{\mu\nu},B]\rangle + b_{13}\langle\overline B \sigma_{\mu\nu} \{f_+^{\mu\nu},B\}\rangle \Big),
\end{align}
where $f_+^{\mu\nu}$ includes the electromagnetic field strength tensor, which vanishes for $v^\mu=0$ but will become important for the photoproduction amplitude later. All operators of the second chiral order in Eq.~\reff{eqn:LAGR1} are accompanied by the low-energy constants (LECs) $b_i$. We complete the driving term Eq.~\reff{VWT} of Eq.~\reff{eqn:BSE} by the second order chiral potential, which reads
\begin{align}\label{VNLO}
 V_{NLO}(q_2, q_1; p)=&A_{14}(q_1\cdot q_2)+A_{57}[\slashed{q_1},\slashed{q_2}]+A_{M}  \nn\\
 &+A_{811}\big(\slashed{q_2}(q_1\cdot p)+\slashed{q_1}(q_2\cdot p)\big),
\end{align}
where the matrices $A_{...}$ in the channel space depend on the LECs as detailed in App.~\ref{app:coupling}. As a matter of fact the importance of the second order terms in the kernel of the BSE is twofold. First of all, as can be seen in \cite{Mai:2009ce}, such terms lead to sizable corrections of the meson-baryon scattering amplitudes. Secondly, the contact interactions of the second chiral order not only contribute to the s-wave but also to the p-waves, which are then iterated in the BSE. In App.~\ref{app:pwa} we demonstrate in a toy model that the presence of the first two partial waves is sufficient to reproduce the correct behavior of the differential cross sections at sufficiently low energies.

\subsection{Solution of the BSE}

For the solution of the BSE we use the recipe developed in a foregoing
publication \cite{Bruns:2010sv}. As described there we utilize dimensional
regularization to treat the divergent loop integrals, where the purely
baryonic integrals are set to zero from the beginning, while only an
energy-independent constant is subtracted from the fundamental meson-baryon
loop integral. This prescription to treat the large baryon mass scale is
similar to the EOMS regularization scheme described in \cite{Fuchs:2003qc}. On
the other hand it resolves several technical problems, which appear in the
course of the study of the photoproduction amplitudes described in the next
section. To further extent, the solution of the BSE corresponds to an infinite
chain of one-meson-one-baryon loop diagrams, see Fig. \ref{pic:BSE}. From the
point of view of the usual  perturbative treatment this would demand an
infinite number of counterterms from a local Lagrangian to absorb the loop
divergences. This is of course not feasible in an effective field theory. In
our non-perturbative framework, the ignorance of higher-order terms in the
scattering kernel, which would serve to cancel the divergences and the
scale-dependence of the loop integrals in a perturbative setting, reflects
itself by the appearance of a new free parameter for every loop-integration,
parametrized here by the logarithm of the renormalization scale. This
pragmatic approach is commonly adopted in the literature, see
e.g. \cite{Nieves:1999bx,Nieves:2001wt, Bruns:2010sv,Mai:2012dt,
  Ikeda:2012au}. The new free parameters are not completely arbitrary,
however: At least, we must impose that the values for the renormalization
scale correspond to neglected higher order terms of natural size. Should this
not be the case, and a scale of e.g. $\mu\sim\mathrm{TeV}$ emerge from some
fits, we must discard that solution as unnatural. As a side remark, we note
that any modification of the loop integrals corresponds to a specific
modification of the potential $V$ in the solution of the BSE. For an explicit
demonstration of this procedure we refer to App.~F of \cite{PCB:Diss}. The
requirement that the modification of the potential is not dominating the
leading order terms also yields the mentioned constraints on the free
scales. In conclusion, the foregoing discussion suggests that it is sufficient
in the present work to apply the subtraction scheme described above, keeping
in mind that the modified loop integrals still depend on the renormalization
scale, which constitutes a free fitting parameter. In the next subsection we
will re-examine this as well as the possibility to adjust this scale to a
fixed value  due to constraints on the loop dressing of vertex functions.

The essential advantage of the above treatment is the preservation of the analytic structure of the loop integrals, which allows for a continuation of the scattering amplitudes into the complex energy plane. The solution of the BSE is presented in App.~\ref{app:solBSE}. It can be written in terms of elementary functions (that is without resorting to a numerical solution) of the loop integrals, which are collected in App.~\ref{app:loop}. Once the BSE has been solved we put the external particles on their mass shell and calculate partial wave amplitudes as well as differential cross section for hadronic scattering. For the evaluation of the photoproduction amplitudes we will require the full off-shell dependence of the hadronic solution, as will be described later.

\subsection{Fit}

It is important to clarify the physical input to the scattering
amplitudes. Throughout this work we will use physical hadron masses (in GeV),
i.e. $M_{\pi^0}=0.1350$, $M_{\pi^+}=0.1396$, $M_\eta=0.5478$,
$M_{K^+}=0.4937$, $M_{K^0}=0.4977$, $m_p=0.9383$, $m_n=0.9396$,
$m_\Lambda=1.1157$, $m_{\Sigma^0}=1.1926$ and $m_{\Sigma^+}=1.1894$. The
baryon mass in the chiral limit, $m_0$ in Eq.~\reff{eqn:LAGR1}, can be fixed
to $1$~GeV without loss of generality, as any other value only amounts to a
rescaling  of the unknown LECs. In contrast to the earlier work
\cite{Borasoy:2007ku}, the meson decay constants are fixed to their physical 
values, i.e. $F_\pi=F_\eta/1.3=0.0924$~GeV, $~F_K =0.113$~GeV.

To pin down the free parameters of our approach we have to specify experimental input available on the market for the considered meson-baryon channels. From the experimental point of view elastic $\pi N$ scattering is by far the best explored reaction. On the other hand it is clear that the low-energy region is dominated by the p-wave resonances, namely Roper and Delta. Our foregoing investigations in \cite{Bruns:2010sv} have shown that we are not able to dynamically generate these resonances consistently with the s-wave resonances. Since those degrees of freedom are not included in our approach, we restrict ourselves to the analysis of s-waves. We fit our results for these $\pi N$ partial waves to the widely accepted partial wave analysis (WI08) by the SAID collaboration \cite{Workman:2012hx}. Comparing an earlier analysis by the Karlsruhe group \cite{Koch:1985bn} to the current one, we assign for the energies below $W=1.28$~GeV an absolute systematic error of $0.005$ and for higher energies an error of $0.030$ to the partial wave amplitudes. To some extent this is in agreement with error estimates done in \cite{Nieves:2001wt}, which are motivated by the expectation of pronounced three-body effects above the $\pi\pi N$ threshold.

Another widely explored channel is $\pi^- p \rightarrow \eta n$, for which we
consider quite recent but already very established results on differential
cross sections measured by Prakhov et al. in \cite{Prakhov:2005qb}. For all
seven measured incident pion beam momenta $p_{\rm lab}$ we assign a
measurement error as well as the systematic error of 6$\%$ as pointed out in
Ref.~\cite{Prakhov:2005qb}. Moreover one should keep in mind that also $p_{\rm
  lab}$ itself entails an uncertainty, which hampers the clear pairwise
separation between most of the given beam momenta \cite{emailPrakov}. 
We do not consider this uncertainty in our fitting routine as the inclusion
would require an additional model-dependent input. The necessary formalism is collected in App.~\ref{app:pwa}.

To fit the above data we follow two different fit strategies. \textbf{Fit
  strategy (I)}: We start from the best fit obtained in \cite{Bruns:2010sv}
and additionally include the $\pi^- p \rightarrow \eta n$ differential cross
sections by Prakhov et al. adjusting all 17 parameters of the   model
($\log(\mu_\pi), \log(\mu_K), \log(\mu_{\eta})$ as well as $14$ LECs). The
fitting region in the elastic $\pi N$ channel is chosen to be
$(m_p+M_\pi)<W_{{\rm cms}}<1.56$ GeV for both $S_{11}$ and $S_{31}$. It is
obvious that the new data will restrict our parameter set additionally,
possibly corrupting the agreement of the elastic $\pi N$ to the SAID data
compared to the fit obtained in  \cite{Bruns:2010sv}. \textbf{Fit strategy
  (II)}: One of the main observations of \cite{Bruns:2010sv} was that the
$S_{11}$ but not the $S_{31}$ partial wave of $\pi N$ scattering can be
described well in the resonance region. On the other hand the main goal of the
present work is to see how hadronic resonances manifest themselves in the
photoproduction amplitude. Moreover since the $\eta N$ final state is an
isospin $1/2$ state we fit the elastic $S_{11}$ partial wave in the energy
region $(m_p+M_\pi)<W_{{\rm cms}}<1.7$ GeV together with $\pi^- p \rightarrow
\eta n$ differential cross sections by Prakhov et al.. The $S_{31}$ is
considered only in the near-threshold region for $W_{{\rm cms}}<1.2$
GeV. Furthermore,  we reduce the number of the free parameters of the model to LECs fixing the regularization scales from the beginning. It turns out that the corrections to the tree-level result of the photoproduction multipole $E_{0+}$ due to the dressing of the $B\rightarrow \phi B$ vertex are large already at the $\pi N$ threshold. Thus we choose the scales such that the meson-baryon loop integral evaluated at $s=m_p^2$ vanishes in every meson-baryon channel,  to assure that the axial vertex-function does not deviate much from the corresponding tree-level expression. 

For both fit strategies we minimize the following quantity
\begin{align}\label{eqn:chi}
 \chi^2:=\chi^2_{DOF}=\frac{\sum_i n_i}{N(\sum_i n_i-p)}\sum_i\frac{\chi^2_i}{n_i}.
\end{align}
Here $p$ is the number of free parameters, $n_i$ is the number of data points available for the observable $i$ and $N$ is the number of observables. This choice of $\chi^2$ ensures the equal weight of both fitted observables, compensating for the different number of data points. Albeit the solution of the BSE is fully analytical, it costs a huge amount of computational power. Thus the fitting as well as error estimation routine is performed in a parallelized version on 20-30 threads of the HISKP cluster utilizing the (migrad) minimization routine of the MINUIT \texttt{C++} library \cite{MINUIT}. The uncertainty of the model is estimated as follows. First, after obtaining the best fit ($\chi^2_{BEST}$) the errors on the parameters are calculated in the (hesse) subroutine of the MINUIT package. Then within these errors we generate a large number of parameter sets ($\sim 10,000$) and calculate for each the corresponding $\chi^2_{DOF}$. Then each set corresponding to a $\chi^2<\chi^2_{BEST}+1.15$ is considered to produce results in the $1\sigma$ region around the central value\footnote{One might argue, whether or not one should divide by the number of degrees of freedom. It turns out that the error bars do not change significantly.}.

\subsection{Results}

\begin{figure*}[tbp]
\includegraphics[width=0.8\linewidth]{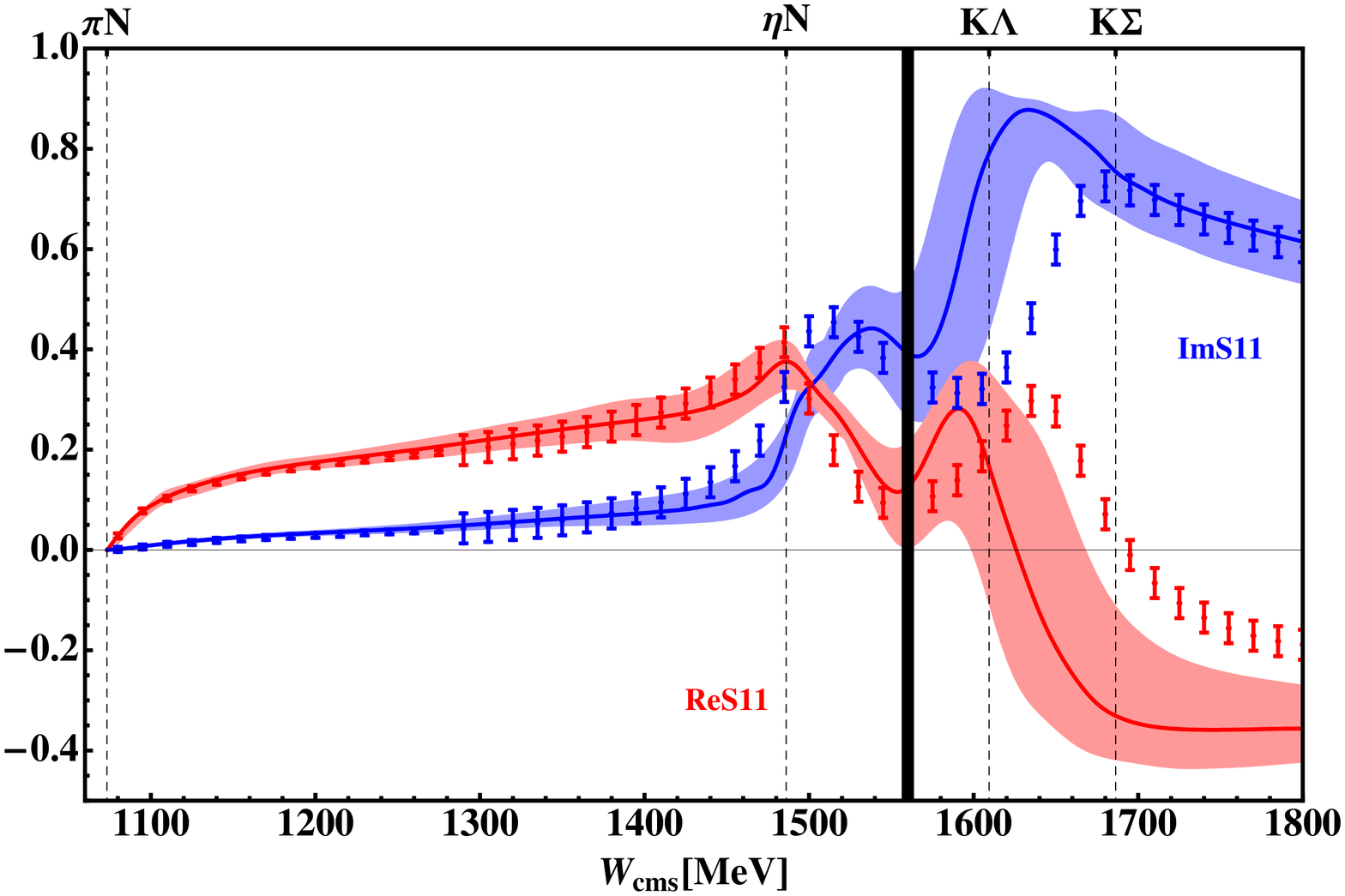}
\caption{Best fit according to fit strategy (I) to the real and imaginary part of the $S_{11}$ partial wave compared to the WI08 analysis done by the SAID collaboration \cite{Workman:2012hx}. The dashed vertical lines correspond to the two particle thresholds and the bold vertical line limits the energy range, up to which the fit has been performed. The blue and red bands represent the 1$\sigma$ uncertainty of our approach as described in the main text.}\label{pic:RES1}
\rule{1.0\textwidth}{1pt}
\includegraphics[width=0.8\linewidth]{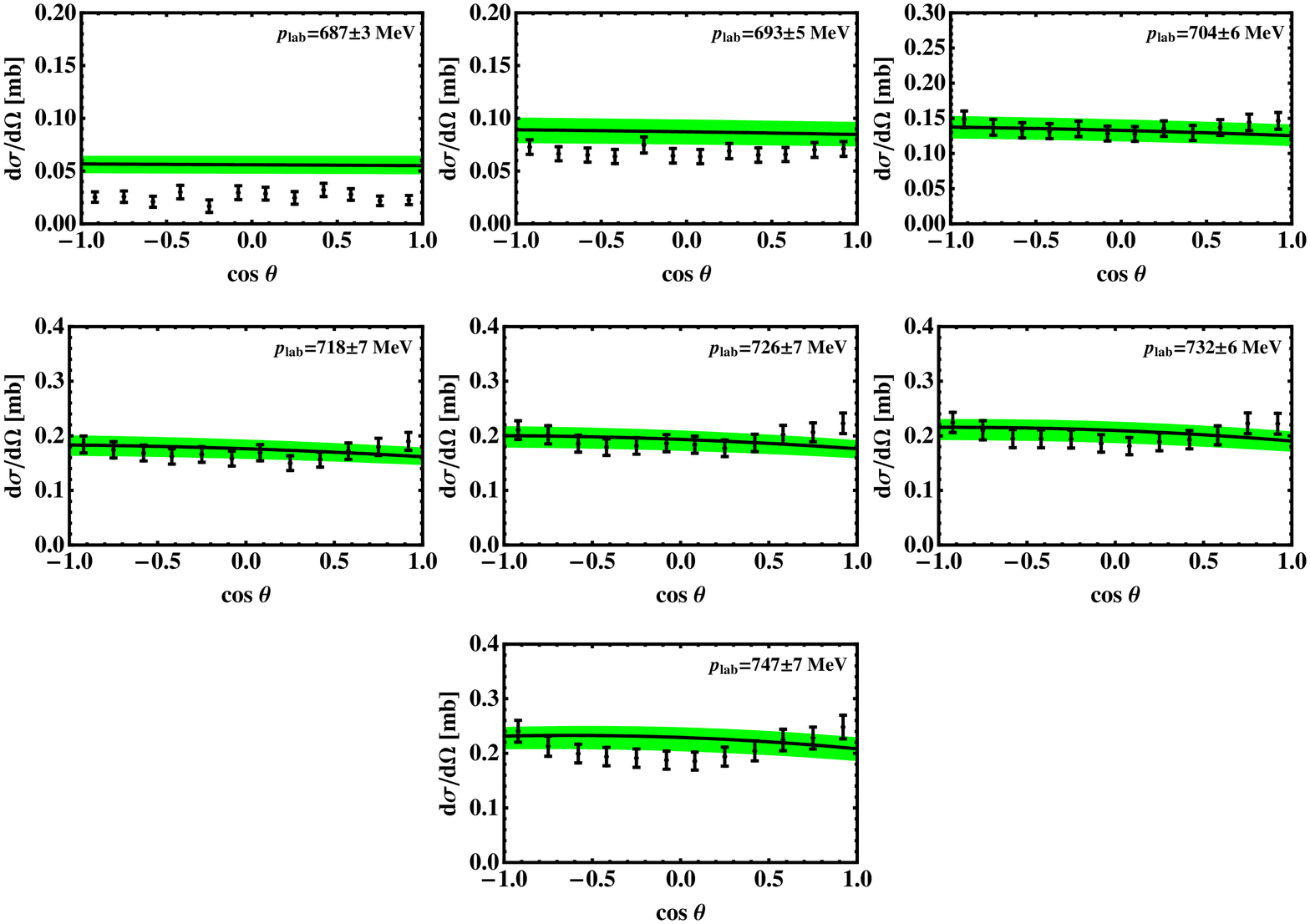}
\caption{Best fit according to fit strategy (I) to the differential cross
  sections for $\pi^-p\to \eta n$ from Ref.~\cite{Prakhov:2005qb}. The error bars of experimental
  data include the systematic error of 6$\%$ as argued in
  Ref.~\cite{Prakhov:2005qb}. 
 The green bands represent the 1$\sigma$ uncertainty of our approach as described in the main text.}\label{pic:RES2}
\end{figure*}

\textbf{Solution I:} Following the first fit strategy, we obtain the best fit
as presented in Figs~\ref{pic:RES1},~\ref{pic:RES2}. 
As already discussed, the differential cross sections on $\pi^- p \to
\eta n$ are observed at seven different pion beam energies, which by
themselves entail a non-negligible uncertainty. 
The latter, however is not included into the definition of the $\chi^2$ for
the reasons given above. 
Therefore, we refrain from giving any numerical value for this quantity\footnote{We wish to note that $\chi^2$ restricted to the SAID data lies only slightly above the one given in the foregoing publication \cite{Bruns:2010sv}, where no other than elastic $\pi N$ channels were included as observables.}. The corresponding parameters ($b_i$ in GeV$^{-1}$ and $\mu_{i}$ in GeV) are of natural size and read:
\begin{center}
\begin{tabular}{c c c c}
\hline
~~~~~~~~~&$\log(\mu_\pi/(1{\rm GeV}))	$&$=+1.003\pm  0.331$&~~~~~~~~~\\
~~~~&$\log(\mu_\eta/(1{\rm GeV})) 	$&$=+1.034\pm  0.298$&~~~~\\
~~~~&$\log(\mu_K/(1{\rm GeV}))		$&$=-0.168\pm  0.080$&~~~~\\
\end{tabular}
\begin{tabular}{c | c}
\hline
~~$b_{1~}=-0.126\pm  0.039$~~		&~~$b_{8~}=+0.610\pm  0.012$~~\\
~~$b_{2~}=-0.139\pm  0.045$~~		&~~$b_{9~}=-0.677\pm  0.037$~~\\
~~$b_{3~}=-2.227\pm  0.133$~~		&~~$b_{10}=+2.027\pm  0.100$~~\\
~~$b_{4~}=-0.288\pm  0.080$~~		&~~$b_{11}=-0.847\pm  0.027$~~\\
~~$b_{5~}=-1.402\pm  0.094$~~		&~~$b_{0~}=-1.063\pm  0.038$~~\\
~~$b_{6~}=+0.474\pm  0.118$~~		&~~$b_{D~}=+0.771\pm  0.042$~~\\
~~$b_{7~}=-1.751\pm  0.368$~~		&~~$b_{F~}=-0.169\pm  0.054$~~\\
\hline
\end{tabular}
\end{center}

\begin{figure}[t]
\begin{center}
\includegraphics[width=1\linewidth]{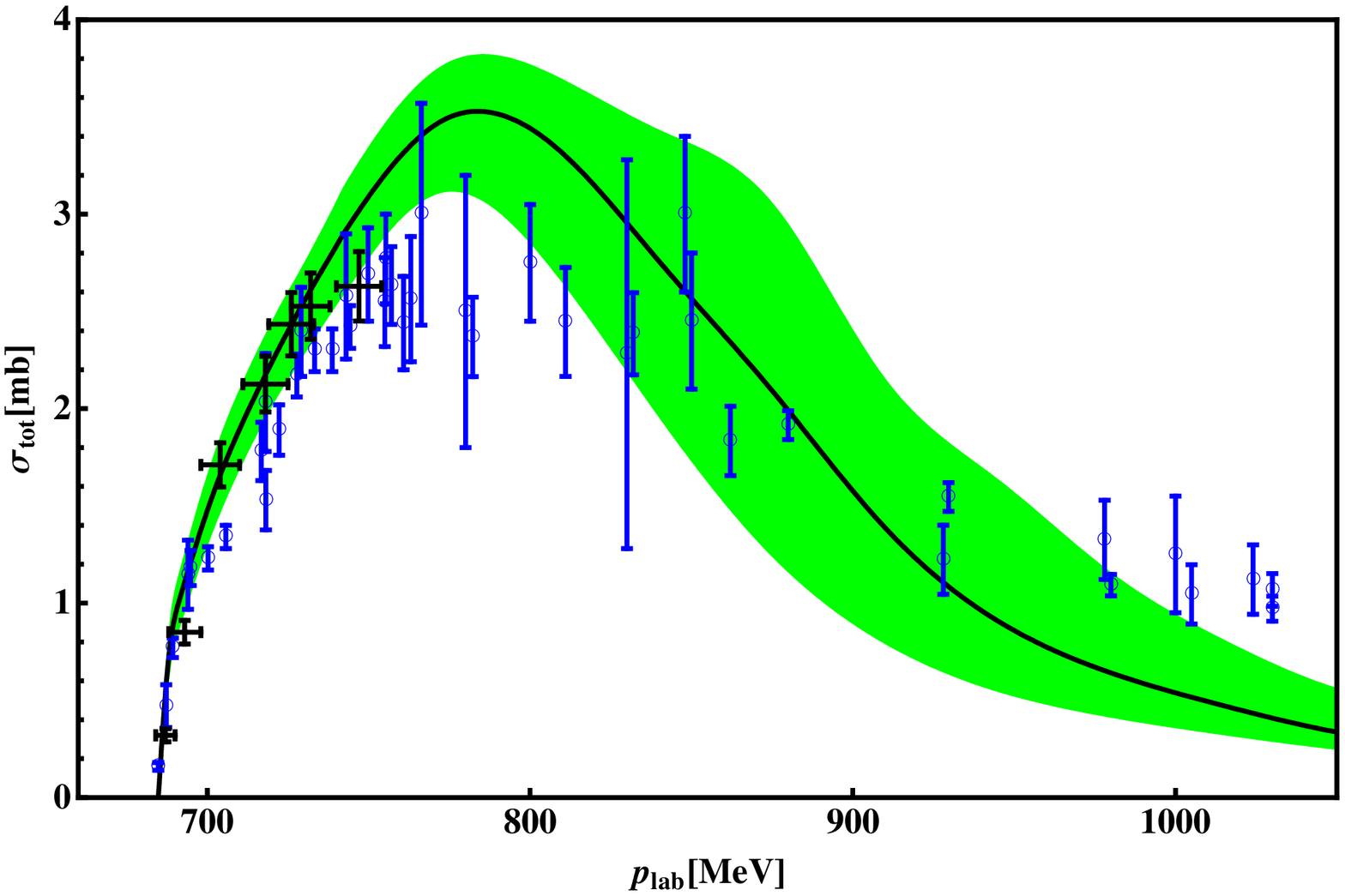}
\caption{Total cross section of the best fit  for $\pi^-p\to \eta n$ according to fit strategy
  (I). The model is fixed to reproduce the differential cross section and thus
  the total cross section as measured by Prakhov et al. (black symbols). The black curve including the uncertainty 
  band is the result of our model. The blue circles  correspond to 
  older measurements as selected by the SAID collaboration which are presented for completeness.}\label{pic:RES3}
\end{center}
\end{figure}

The observation in the elastic $\pi N$ channels is similar the one made in \cite{Bruns:2010sv}. Between the $\pi N$ and $\eta N$ thresholds both partial waves can be fitted nicely to the SAID partial wave analysis. For the $\pi N$ scattering lengths of isospin $I$, $a_{I}$ (in units of $10^{-3}/M_{\pi^+}$), we obtain:
\begin{align}
 a_{3/2}^{\pi N} = -87.0^{+4.3}_{-4.2} \quad \text{and}\quad a_{1/2}^{\pi N} = +174.5^{+15.2}_{-32.8}.
\end{align}
The theoretically cleanest determination of these observables stems from the
analysis of pionic hydrogen and pionic deuterium data based on effective field
theory~\cite{Baru:2010xn}, $a_{1/2} = (179.9\pm 3.6) \times 10^{-3}/M_{\pi^+}$
and $a_{3/2} = (-78.5\pm 3.2) \times 10^{-3}/M_{\pi^+}$, which is in nice
agreement with our determination for the $I=1/2$ channel, but is 
slightly too small for $I=3/2$. For both isospins our determination agrees perfectly with those from the direct extraction by the SAID collaboration: $a_{1/2} = (174.7\pm 2.2) \times 10^{-3}/M_{\pi^+}$ and $a_{3/2} = (-89.4\pm 1.7) \times 10^{-3}/M_{\pi^+}$.\footnote{We thank Ron Workman for providing us with these values.} 

\begin{figure}[tb]
\begin{center}
\includegraphics[width=0.8\linewidth]{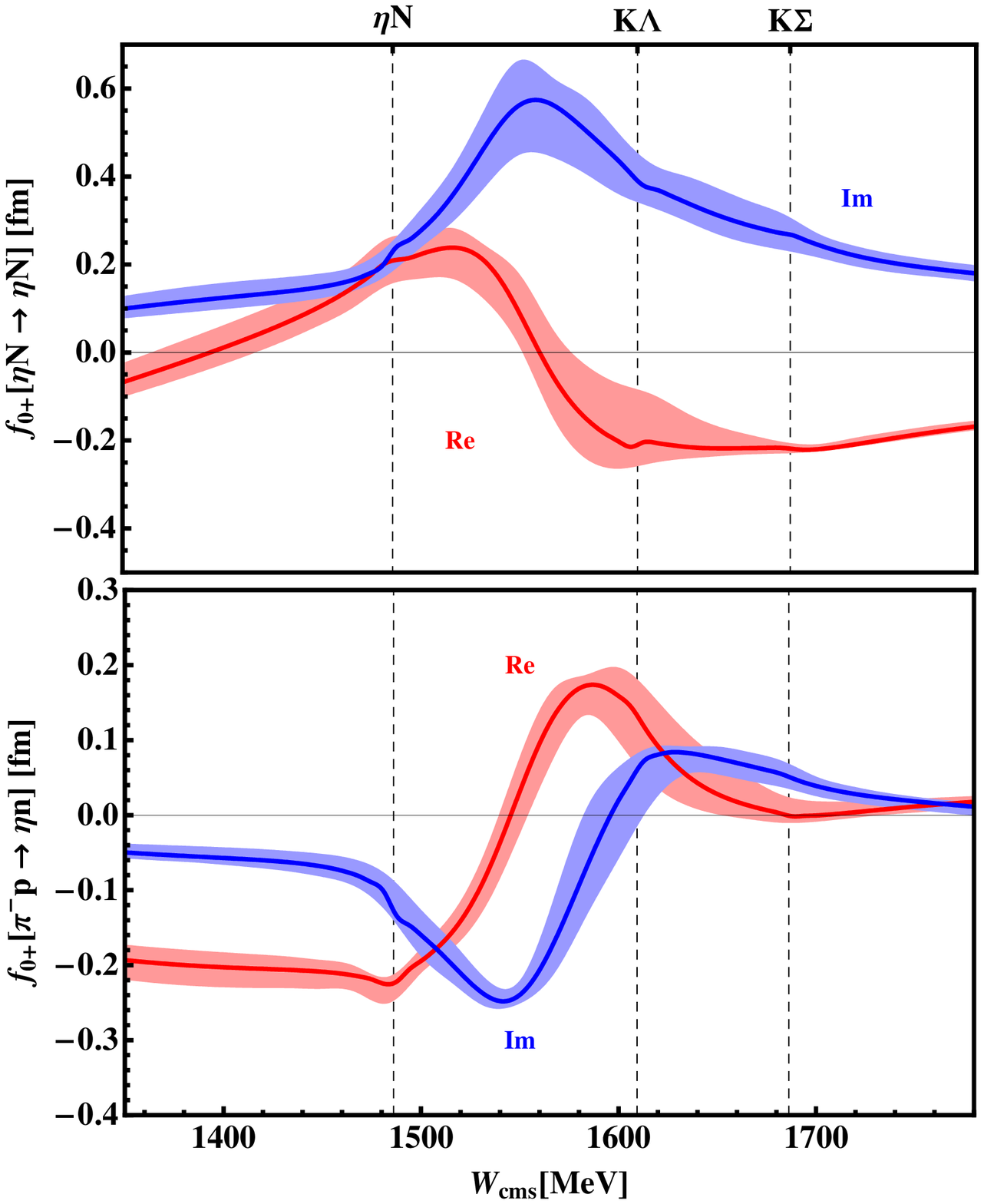}
\caption{The result of our model in the fit strategy (I) for the real and
  imaginary part of the s-wave scattering amplitude of the $\eta N$ (top) and 
  $\pi^- p\to \eta n$ (bottom) channels. The error bands represent the
  uncertainty due to the variation of model parameters as described in the text.}\label{pic:RES4}
\end{center}
\end{figure}

In the higher energy region the lowest $S_{11}$ but not $S_{31}$ resonances could be reproduced as dynamically generated states in our model. The pole position can be extracted via analytic continuation of the scattering amplitude to the complex $W_{{\rm cms}}$ plane and read
\begin{align}\label{polesI}
	W_{1535}&= (1.547^{+0.004}_{-0.021} - i 0.046^{+0.004}_{-0.017} ) \text{ GeV}, \nn \\
	W_{1650}&= (1.597^{+0.017}_{-0.020} - i 0.045^{+0.010}_{-0.015} ) \text{ GeV}.
\end{align}
Obviously the inclusion of $\eta N$ observables into the analysis forces the pole of $N^*(1535)$ to the higher and the pole of $N^*(1650)$ to the lower energies compared with the previous analysis \cite{Bruns:2010sv}. This observation is in some agreement with the analysis in \cite{Gasparyan:2003fp}. There in a meson-exchange model the analysis of the inelasticities has shown that a simultaneous description of the $\eta N$ and $\pi N$ scattering amplitude is hampered by the missing $\pi\pi N$ channels, which are also missing in our approach. However let us repeat that the starting values of the present fit strategy are chosen to be those from \cite{Bruns:2010sv}. Although there is no reason to doubt about them for the elastic $\pi N$ scattering one should keep in mind that an inclusion of additional, i.e. $\eta N$ data alter the amplitudes in the $\pi N$ channel as well.

For the pion induced eta production, Fig.~\ref{pic:RES2}, we observe that the outcome of the model agrees with the experimental data, keeping in mind the uncertainty on the pion beam momenta. The inclusion of the latter is crucial especially for the lowest pion beam momenta, where the slope of the total cross section is enormous as can be seen in Fig. \ref{pic:RES3}. There we present the outcome of the model for higher beam momenta than included in the fit. Obviously the total cross section agrees with the experimental data within the error bars. We also observe a large qualitative agreement of the outcome of the model with the older and less precise measurements, selected by the SAID collaboration.
The overshooting of the total $\pi N \rightarrow \eta N$ cross section by $30\%$ as discussed in \cite{Gasparyan:2003fp} is overcome obviously via moving of the $N^*(1535)$ peak towards higher energies, which is a direct consequence of the present fit strategy. In App.~\ref{app:pwa} we show that the present model is in principle capable to simulate a $\cos^2 \theta$ like behavior in the differential cross sections, usually referred to as the influence of the $D_{13}$ resonance. The observation to be made from Fig. \ref{pic:RES2} is that the inclusion of the elastic $\pi N$ channels prevents (or at least damps) such behavior. One should notice that the curvature in the data is of comparable size as the uncertainty in the data as well as the uncertainty band of our approach.

We can make a prediction of the s-wave amplitudes for elastic $\eta N$ as well as $\pi^- p\rightarrow \eta n$ scattering as presented in Fig. \ref{pic:RES4}. Here and in the future we use the H\"ohler partial waves denoted by $f_{0+}$ in contrast to the $S_{11}$ partial wave used by SAID collaboration, see e.g. Fig. \ref{pic:RES1}, which is the $f_{0+}$ for isospin 1/2 multiplied by $q_{\rm cms}$. In both channels the real and imaginary part shows a similar behavior to the one shown in \cite{Lutz:2005jd}. However the position of the $S_{11}$ peak is systematically shifted to higher energies in our approach, which is again a feature of the present fit strategy. For the scattering lengths we obtain the following values
\begin{align}
 a_{\eta N \rightarrow \eta N}~   &= (+0.219_{-0.068}^{+0.047}  +  i 0.235_{-0.055}^{+0.148}) \text{ fm},  \nn\\
 a_{\pi^- p \rightarrow \eta n}   &= (-0.234_{-0.024}^{+0.020}  -  i 0.129_{-0.104}^{+0.048}) \text{ fm}.
\end{align}
There is a large spread in the results on $\eta N$ scattering lengths debated for a long time, see \cite{Arndt:2005dg} for a nice collection of those. One can note that most models predict a positive real and imaginary part of the scattering length, which is in agreement with our result as well.

\medskip

\begin{figure*}[hpb]
\includegraphics[width=0.8\linewidth,height=\textheight, keepaspectratio]{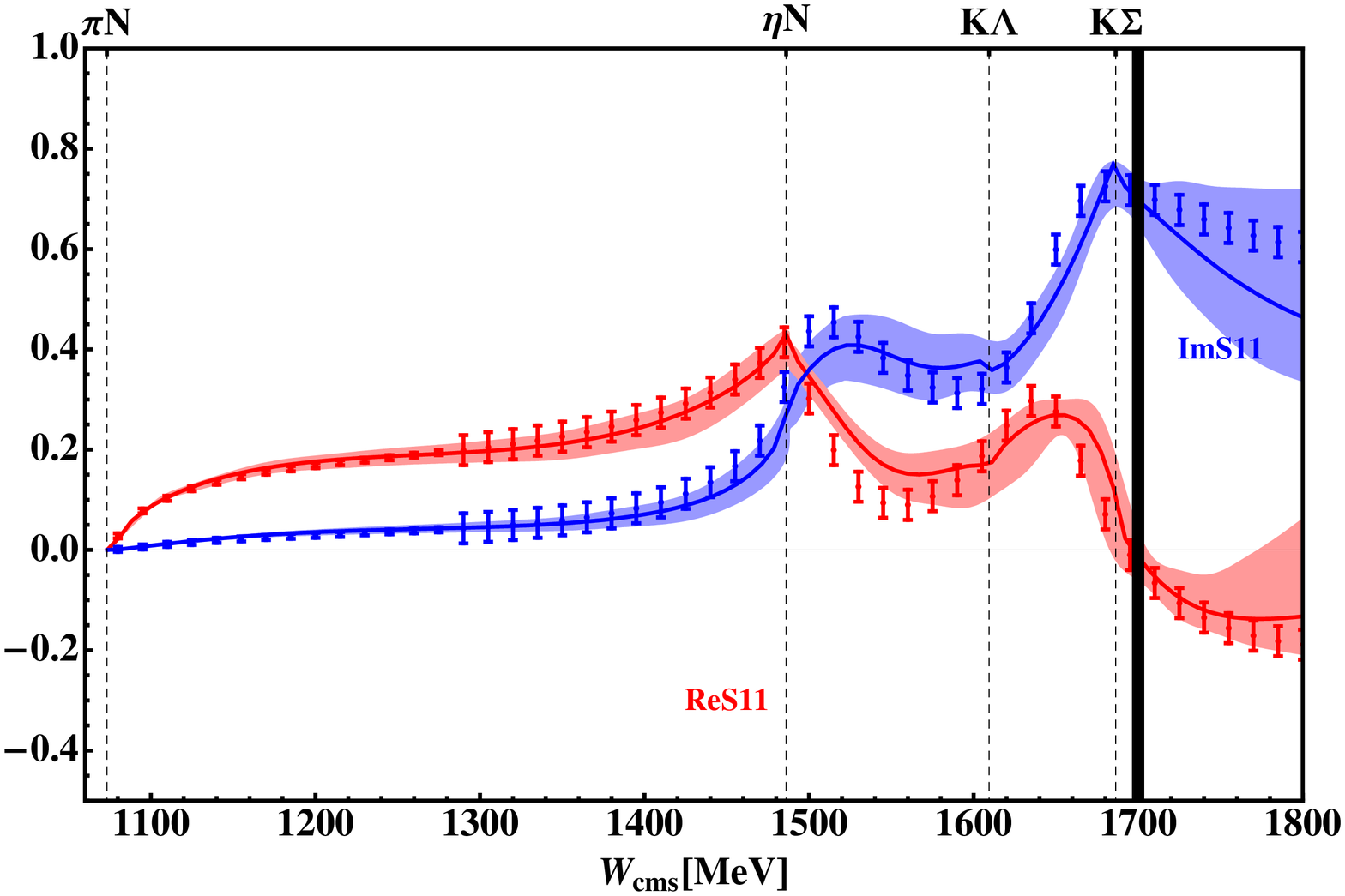}
\caption{Best fit according to fit strategy II to the real and imaginary part of the $S_{11}$ partial wave compared to the WI08 analysis done by the SAID collaboration \cite{Workman:2012hx}. The dashed vertical lines correspond to the two particle thresholds and the bold vertical line limits the energy range, up to which the fit has been performed. The blue and red bands represent the 1$\sigma$ uncertainty of our approach as described in the text.}\label{pic:RES1new}
\rule{1.0\textwidth}{1pt}
\includegraphics[width=0.8\linewidth,height=\textheight, keepaspectratio]{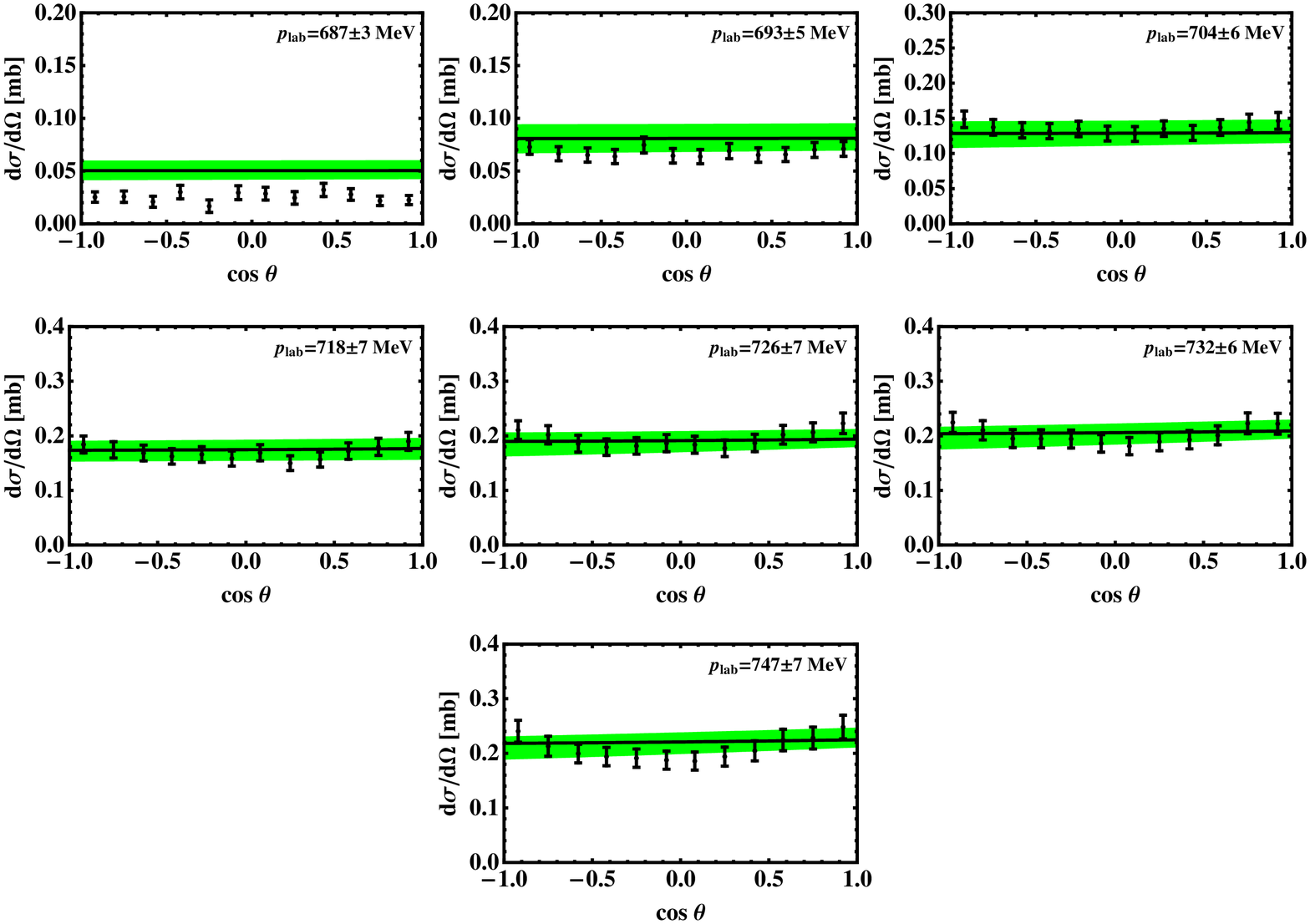}
\caption{Best fit according to fit strategy II to the differential cross sections for $\pi^-p\to \eta n$ from Ref.~\cite{Prakhov:2005qb}. The error bars of experimental data include the systematic error of 6$\%$ as argued in Ref.~\cite{Prakhov:2005qb}. The green bands represent the 1$\sigma$ uncertainty of our approach as described in the text.}\label{pic:RES2new}
\end{figure*}

\textbf{Solution II:} As argued above, the number of free parameters is reduced in this fit strategy by the three regularization scales. They are fixed such that for each meson-baryon channel ($i$): $I_{MB}(m_p^2,m_i,M_i)\overset{!}{=}0$ in the nomenclature of App.~\ref{app:loop}, which yields the following values
\begin{align*}
  \log(\mu_\pi)=-0.368, \quad \log(\mu_\eta)=0.056, \quad \log(\mu_K)=0.210\,.          
\end{align*}
The best fit of the 14 LECs, which are the only free parameters of the model in the present fit strategy, is presented in Fig. \ref{pic:RES1new} and \ref{pic:RES2new}. All parameters are of perfect natural size and read including the error bars (in GeV$^{-1}$):
\begin{center}
\begin{tabular}{c | c}
\hline
~~$b_{1~}=-0.014\pm  0.023$~~		&~~$b_{8~}=+0.272\pm  0.015$~~\\
~~$b_{2~}=-0.207\pm  0.051$~~		&~~$b_{9~}=-0.483\pm  0.032$~~\\
~~$b_{3~}=-1.063\pm  0.032$~~		&~~$b_{10}=+1.054\pm  0.021$~~\\
~~$b_{4~}=-1.312\pm  0.023$~~		&~~$b_{11}=+0.328\pm  0.015$~~\\
~~$b_{5~}=-0.628\pm  0.060$~~		&~~$b_{0~}=-1.228\pm  0.005$~~\\
~~$b_{6~}=+0.508\pm  0.045$~~		&~~$b_{D~}=+1.097\pm  0.011$~~\\
~~$b_{7~}=+1.041\pm  0.191$~~		&~~$b_{F~}=-0.858\pm  0.011$~~\\
\hline
\end{tabular}
\end{center}
In the elastic $\pi N$ channel the $S_{11}$ partial wave agrees almost perfectly in the whole energy range with the one from the analysis by the SAID collaboration. The corresponding scattering lengths are extracted to be (in units of $10^{-3}/M_{\pi^+}$)
\begin{align}
 a_{3/2}^{\pi N} = -93.0^{+4.7}_{-6.3} \quad \text{and}\quad a_{1/2}^{\pi N} = +168.9^{+5.9}_{-6.4}.
\end{align}
A comparison with the result of other calculations, given before, shows the
same pattern as in the previous fit. Both scattering lengths agree within the
error bars with the direct extraction by the SAID collaboration and are
smaller than the values extracted in Ref.~\cite{Baru:2010xn}.

Both $N^*(1535)$ and $N^*(1650)$ are reproduced as dynamically generated states of the lowest meson and baryon octet states. The pole positions of both $N^*$ resonances read
\begin{align}\label{polesII}
	W_{1535}&= (1.512^{~+8}_{~-7} - i 0.070^{~+9}_{~-5}) \text{ GeV}, \nn \\
	W_{1650}&= (1.715^{+32}_{-24} - i 0.116^{+15}_{-24}) \text{ GeV}.
\end{align}
As a matter of fact we expect the pole positions from the present fit strategy
to be even more realistic than those from the previous fit strategy as well as
from the analysis done in \cite{Bruns:2010sv}, where no physical information
was included for energies in the region of the second resonance. The pole
position of the $N^*(1535)$ is perfectly within the uncertainty band presented
in \cite{Yao:2006px}, i.e. $W_{1535}= (1.490...1.530)- 
i (0.045...0.125)$ GeV. On the other hand the position of $N^*(1650)$ differs slightly from the one given there, i.e. $W_{1650}= (1.640...1.670)- i(0.050...0.085)$ GeV. Note that both bands in \cite{Yao:2006px} are mostly based on a selection of partial wave analyses. The pole positions from two comparable theoretical works read 
$W_{1535}=1.519-i0.064$~GeV and $W_{1650}=1.669-i0.068$~GeV from \cite{Doring:2009uc} as well as $W_{1535}=1.496-i0.041$~GeV and $W_{1650}=1.686-i0.096$~GeV from \cite{Nieves:1999bx}. 

\begin{figure}[ptb]
\includegraphics[width=1\linewidth]{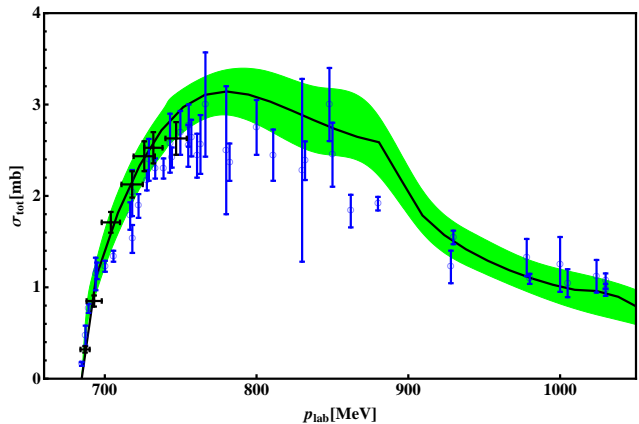}
\caption{Total cross sections of the best fit  for $\pi^-p\to \eta n$
 according to fit strategy (II). The model is fixed to reproduce the differential cross section and thus also the total cross section as measured by Prakhov et al. (black symbol). The black curve including the uncertainty band is the outcome of the model. The blue circles correspond to the older measurements as selected by the SAID collaboration which are only presented for completeness.}\label{pic:RES3new}
\end{figure}

For the pion induced eta production we observe that,  taking into account the
uncertainty of the pion beam energy, all seven differential cross sections
agree with the data by Prakhov et al., see Fig.~\ref{pic:RES2new}. Again the
$\cos^{2} \theta$ behavior does not appear. 
We have discussed  in App.~\ref{app:pwa} that in principle such a behavior
could be reproduced in our amplitude 
by means of enhanced contributions from the p-waves, which are iterated in our
approach. We conclude from this observation that 
such a behavior is excluded in this combined $\pi N$ and $\eta N$ fit. In
Fig.~\ref{pic:RES3new} we present the 
total cross section for the same process beyond the fitting region. In
contrast to the previous fit, 
we observe here a behavior of the resulting cross section $\sigma(p_{\rm
  lab})$ much more in line with our 
earlier analysis\footnote{That means that no forced shift of the $N^*(1535)$ pole to higher energies emerges from the fits.}. As a matter of fact we do not see any overprediction of the total cross section at the position of the $N^*(1535)$ peak which has been pointed out before, relying on the analysis of \cite{Gasparyan:2003fp}.

\begin{figure}[ptb]
\includegraphics[width=0.8\linewidth]{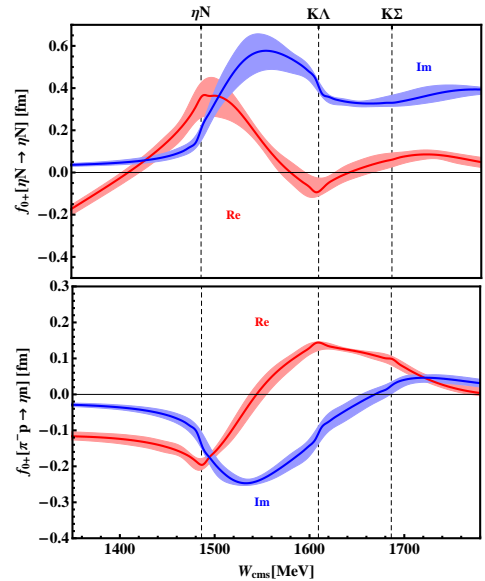}
\caption{The result of the model employing fit strategy (II) for the real and
  imaginary part of the s-wave 
scattering amplitude of the $\eta N$ (top) and $\pi^- p\to \eta n$ (bottom)
channels. The error bands express the uncertainty due to variation of the 
model parameters as described in the text.}\label{pic:RES4new}
\end{figure}

As a further prediction we extract the scattering lengths of the $\eta N$ channels, which read
\begin{align}
 a_{\eta N \rightarrow \eta N}~   &= (+0.378_{-0.101}^{+0.092}  +  i 0.201_{-0.036}^{+0.043}) \text{ fm},  \nn\\
 a_{\pi^- p \rightarrow \eta n}   &= (-0.208_{-0.017}^{+0.016}  -  i 0.138_{-0.029}^{+0.025}) \text{ fm}.
\end{align}
The observation to be made is that both are consistent with the extraction from other (more phenomenological) approaches  \cite{Arndt:2005dg}. The s-wave amplitude in both channels can be found in Fig. \ref{pic:RES4new}. For energies lower than the $K\Lambda$ threshold we observe similar behavior as for the amplitudes extracted in fit strategy I, see Fig. \ref{pic:RES4}.

\subsection{Summary: Hadronic sector}

As an intermediate summary we conclude that the present framework serves as an
appropriate theoretical tool for the analysis of hadronic scattering in a
fairly wide energy range. Both isospin $1/2$ s-wave resonances, the $N^*(1535)$ and $N^*(1650)$ are generated dynamically from the NLO chiral potential in a quite natural way, i.e. resumming the Feynman diagrams in the BSE with the full off-shell dependence. At the same time it describes the pion induced $\eta$ production of the nucleon in full agreement with the experimental data by Prakhov et al..

\clearpage
\begin{widetext}
\begin{figure*}[htb]
\begin{center}
\includegraphics[width=0.8\linewidth]{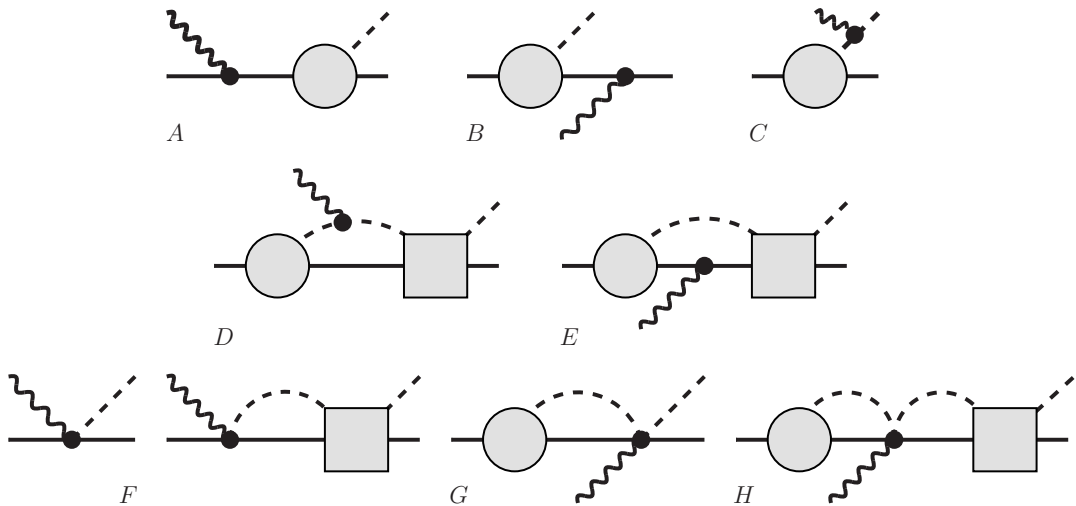}
\end{center}
\caption{Types of diagrams of the turtle approximation. Shaded circles, squares and black dots represent the dressed meson-baryon vertex $\Gamma$, scattering amplitude $T$ and the photon vertex as described in the main text.}\label{pic:photo}
\end{figure*}
\end{widetext}

\section{Pion photoproduction off the proton}\label{sec:Photo}

Pion photo- and electroproduction off protons has been one of the premier
objects of study in hadron physics for decades. One of the major issues was
\cite{Watson:1954uc,Berends:1967vi} and still is the interplay between the
hadronic scattering and the photoproduction of mesons off the baryons. It is
unquestionable that the meson-baryon interaction plays a crucial role in the
photoproduction processes via the rescattering processes. At the production
threshold, pion photoproduction can be successfully analysed within strictly
perturbative ChPT as has been done to one loop 
about twenty years ago \cite{Bernard:1992nc,Bernard:1994gm}.
Going to higher energies, one is again confronted with the problems already
appearing in the hadronic sector as discussed in the previous section, namely,
resonance phenomena. Thus a non-perturbative framework is required to
implement the rescattering mechanism adequately. In the early years the
unitarized hadronic amplitude was simply used as the final state interaction
(FSI) multiplied on top of the $\gamma p \to \pi N$ contact interaction, which in general violates the Ward-Takahashi identities and thus gauge invariance. Recently a framework for pion photoproduction based on the J\"ulich model was constructed in \cite{Huang:2011as}. There the hadronic part of the amplitude is also used as the FSI coupled to a special form of contact term and fulfilling the gauge Ward-Takahashi identities by construction.

Our approach follows a different direction, where gauge invariance is not
enforced via {\em ad hoc\,} conditions on vertex functions and propagators,
but follows most straightforwardly from the selected infinite subset of
Feynman graphs which are summed up. 
The basic ideas can be traced back from
Refs.~\cite{Kvinikhidze:1998xn,Gross:1987bu,vanAntwerpen:1994vh,Borasoy:2005zg}.
Essentially, one adopts a generalization of the construction of a
gauge-invariant amplitude (as spelled out e.g. in sec.~(7.4) of \cite{Peskin})
to the present non-perturbative setting. It was first applied to the analysis
of kaon photoproduction, relying on the leading s-wave terms from the
three-flavor chiral Lagrangian, in \cite{Borasoy:2007ku}. There, the selected
subset of Feynman graphs was referred to as ``turtle approximation''. In
principle this is the most natural way of constructing a gauge invariant
photoproduction amplitude as the photon is coupled to any point of the $p
\rightarrow \phi B$ amplitude, ensuring current conservation. Nevertheless it
requires as an input the underlying hadronic amplitude with the full off-shell
dependence. Such an amplitude is provided in the previous section. It fulfills
the two-body unitarity requirement exactly and the parameters are fixed such
as to reproduce the s-wave of $\pi N$ as well as $\pi N \rightarrow \eta N$
scattering. Without any further fitting we wish to investigate what we can
learn about the multipole amplitudes by just plugging in our fixed hadronic
amplitude as an effective vertex function. To put it in the words of Berends
et al. \cite{Berends:1967vi} ``we wish to see how the resonances from the 
hadronic spectrum manifest themselves in the pion photoproduction observables''.

\subsection{Formalism}

Closely following the formalism explained in \cite{Borasoy:2007ku}, the gauge invariant photoproduction amplitude $\MM(q',k;p)$, is a sum of 8 different types of Feynman diagrams, see Fig.~\ref{pic:photo}. Here, $q'$ is the four-momentum of the produced meson and $p$ is again the overall four-momentum. The four-momentum of the incoming photon is denoted by $k$. The scattering amplitude $T$ has been calculated in the previous section, consequently there are only two buildings blocks left to be clarified, i.e. $\Gamma$ denoting the dressed meson-baryon vertex and the photon vertices $W_{\gamma \phi\rightarrow \phi}$, $W_{\gamma B\rightarrow B}$, $W_{\gamma B\rightarrow B\phi}$ and $W_{\gamma B \phi\rightarrow B \phi}$.

\begin{figure}[t]
\begin{center}
\includegraphics[width=1.0\linewidth]{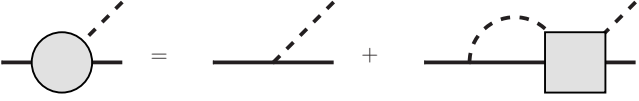}
\end{center}
\caption{Symbolical representation of the dressed meson-baryon amplitude $\Gamma$ (circle), where the shaded square represents the meson-baryon scattering amplitude $T$.}\label{pic:gamma}
\end{figure}

The exact two-body unitarity is a crucial property of the hadronic
amplitude. For this to be preserved in the photoproduction amplitude as well,
the axial meson-baryon coupling has to be dressed properly. The tree level
axial meson-baryon potential stems from the leading order 
chiral Lagrangian Eq.~\reff{eqn:LAGR0} and reads
\begin{align}
 V_{ax}(q')=A_X~\slashed q'\gamma^5,
\end{align}
where we have separated off the channel space structure, which is specified in App.~\ref{app:coupling}. Dressing of this amplitude in the 'turtle approximation' \cite{Borasoy:2007ku} is presented in a rather intuitive pictorial way in Fig. \ref{pic:gamma} and reads
\begin{align}
 \Gamma(q',\tilde p)=&V_{ax}(q') \\
		     &+\int \frac{d^dl}{(2\pi)^d} T(\slashed{q'},\slashed{l};\tilde p)iS(\slashed{\tilde p}-\slashed{l})\Delta(l)V_{ax}(l),\nn
\end{align}
where $\tilde p$ denotes the total four-momentum of this process which can
take values of the proton momentum or the overall four-momentum of the photoproduction process. The scattering amplitude $T$ consists of 20 different Dirac structures as presented in App.~\ref{app:solBSE} and gives rise to 6 different structures of the amplitude $\Gamma$, i.e.
\begin{align}
  \Gamma(q',\tilde p)=~&~\Gamma_1(\tilde p) \cdot \slashed{\tilde p}\gamma^5 + \Gamma_2(\tilde p) \cdot \gamma^5 + \Gamma_3(\tilde p) \cdot \slashed{q'}\slashed{\tilde p}\gamma^5 \nn\\
                      +&~\Gamma_4(\tilde p) \cdot \slashed{q'}\gamma^5 + \Gamma_5(\tilde p) \cdot \gamma^5(q'\cdot\tilde p)  \nn \\
                      +&~\Gamma_6(\tilde p) \cdot \slashed{\tilde p}\gamma^5(q'\cdot\tilde p).
\end{align}
The coefficients $\Gamma_i(\tilde p)$ are elementary functions of $\tilde p^2$, masses, scalar loop integrals, $I_M$ and $I_{MB}(\tilde p^2)$, collected in App~\ref{app:loop}, coefficients $T_i$ of the scattering amplitude as well as of the axial coupling constants $D$ and $F$ from eqn.~\reff{eqn:LAGR0}.

The numerical values of the latter constants should be taken with a grain of salt. Working at tree level for the axial vertex, the sum of $D$ and $F$ is set equal to the value of nucleon axial vector charge $g_A$, which has been in the focus of many experiments for decades and was measured recently to high precision in the neutron $\beta$-decay using ultracold neutrons \cite{Liu:2010ms} to be $g_A=1.27590_{-0.00445}^{+0.00409}$. The ratio $F/D$ is predicted by the $SU(6)_f$ non-relativistic quark model (NRQM) to be $2/3$ which is actually quite close to the value of this ratio extracted from experiments, namely $0.58\pm0.05$, see \cite{Jaffe:1989jz} for a more detailed discussion. Within our approximation the axial coupling enters the photoproduction amplitudes dressed by the meson-baryon loops, for instance the kaon-loops. Those effects are known to be quite sizable and thus one has to choose at which level, i.e. on tree level or that of the dressed vertex $\Gamma$, one wishes to obtain an agreement with the physical (measured) quantities. It turns out that although the value of $\Gamma$ depends strongly on the choice of axial couplings, the photoproduction amplitudes calculated with both sets of axial couplings agree with each other within the uncertainty band. For this reason we stick to the commonly used values of $D=0.8$ and $F=0.5$.


It remains now to specify how to couple the photon to the hadronic skeleton
described above. For consistency reasons we shall consider the photon induced
contact terms up to the second chiral order utilizing the first and second
order chiral Lagrangian, Eq.~\reff{eqn:LAGR0} and 
Eq.~\reff{eqn:LAGR1}. Previously we have set all the external currents to zero except the scalar one, whereas here we consider a vector current $v_\mu=-eQA_\mu$ with the electromagnetic vector potential $A_\mu$ and charge matrix $Q={\rm diag}(2/3,-1/3,-1/3)$. A vector current modifies the covariant derivative as well as the chiral vielbein
\begin{align*}
 [D_\mu, B] =& \partial_\mu B +\frac{1}{2}[\big([u^\dagger,\partial_\mu u]-i(u^\dagger v_\mu u+uv_\mu u^\dagger)\big),B],\\
 u_\mu      =& iu^\dagger \big( \partial_\mu U - i[v_\mu,U]\big) u^\dagger.
\end{align*}
A non-vanishing vector potential also features in $\Lagr^{(2)}_{\phi B}$ via the field-strength tensor 
\begin{align*}
 f_+^{\mu\nu} = u(\partial^\mu v^\nu-\partial^\nu v^\mu)u^\dagger + u^\dagger(\partial^\mu v^\nu-\partial^\nu v^\mu)u.
\end{align*}
It appears in Eq.~\reff{eqn:LAGR1} accompanied by the two LECs $b_{12}$ and
$b_{13}$,
 which can not be determined from the scattering process. As mentioned before
 we do not wish to perform a fit for the photoproduction observables, 
therefore we stick to the values determined in Ref.~\cite{Kubis:2000aa} for these two new LECs. 

Furthermore, the pure mesonic chiral Lagrangian of second chiral order, which reads
\begin{align*}
 \Lagr_\phi^{(2)}=\frac{F_0^2}{4}\langle u_\mu u^\mu + \chi_+\rangle
\end{align*}
gives rise to the photon vertex $W_{\gamma \phi\rightarrow \phi}$. This and the remaining vertices are collected in App.~\ref{app:ver}\,.

\begin{figure*}[t]
\begin{center}
\includegraphics[width=1.0\linewidth]{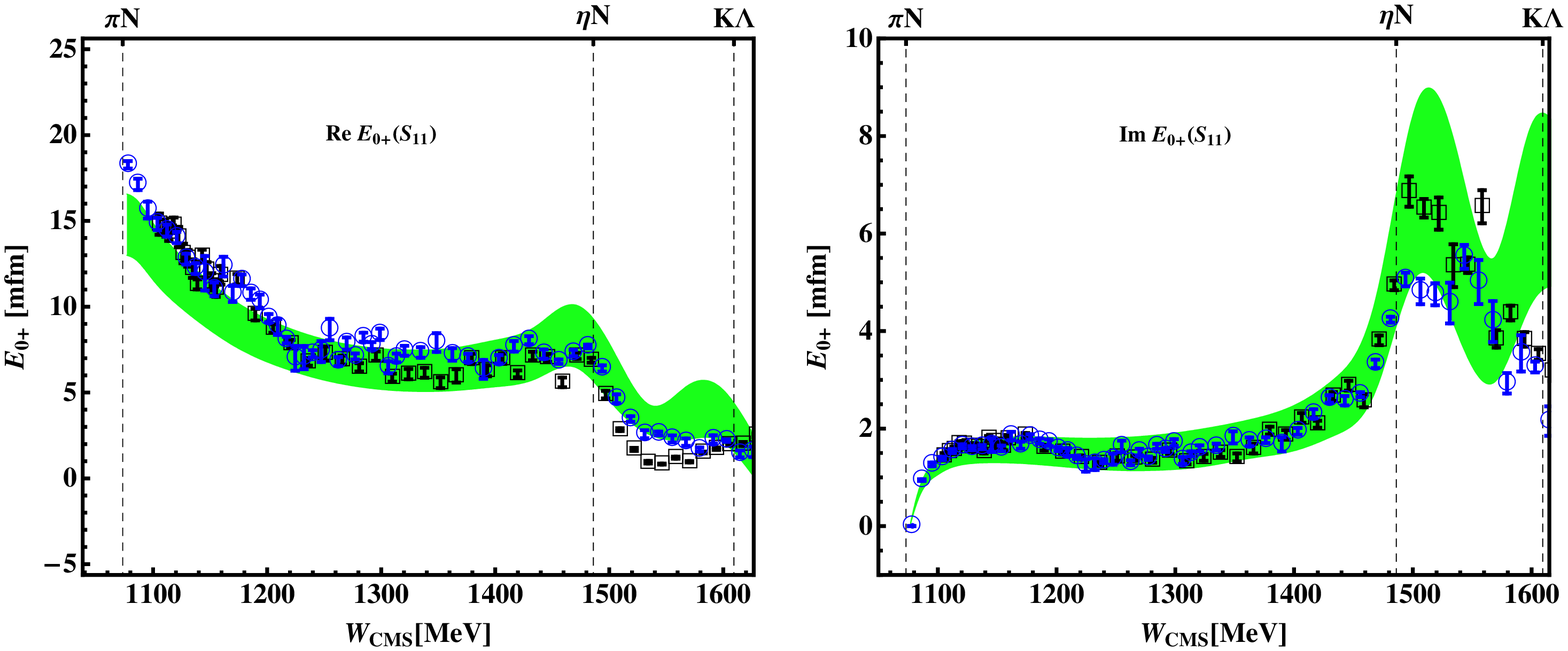}
\caption{Prediction for the multipole $E_{0+}$ for  pion photoproduction
  corresponding to the hadronic solution I. 
For comparison, best fits of the MAID (circles) \cite{Drechsel:2007if} and
SAID (squares) \cite{Workman:2012jf} models 
are represented by blue and black symbols, respectively.}\label{pic:RES5}
\includegraphics[width=1.0\linewidth]{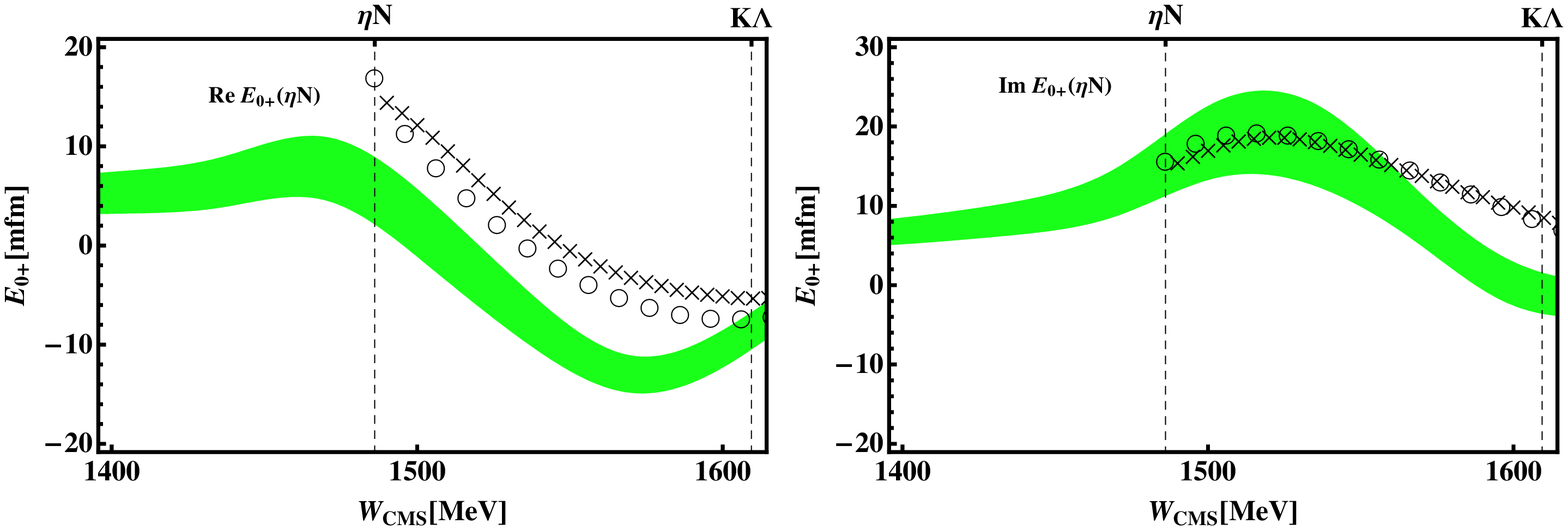}
\caption{$E_{0+}$ for the eta photoproduction as predicted based on hadronic
  solution I. For comparison, we also present the results of  the ETAMAID
  (circles) \cite{Chiang:2001as} 
and Bonn-Gatchina (squares) \cite{Anisovich:2009zy} analyses.}\label{pic:RES6}
\end{center}
\end{figure*}

Having specified all building blocks of the graphs collected in Fig.~\ref{pic:photo} we now calculate the photoproduction amplitude $\mathcal{M}^\mu=\sum_{i=A}^{H} {S_i^\mu}$, where the amplitudes $S_i$ correspond to a respective class of graphs defined in Fig.~\ref{pic:photo}. We wish to emphasize that there are 5 unitarity classes which by themselves obey two-body unitarity: $\{S_A;S_B+S_E;S_C+S_D,S_F;S_G+S_H\}$. Gauge invariance is fulfilled for the amplitudes proportional to $b_{12}$ and $b_{13}$ automatically. On the other hand for the remaining terms it is only fulfilled if all graphs presented in Fig.~\ref{pic:photo} are taken into account, i.e. the photon is coupled to every possible part of the hadronic skeleton.

\begin{figure*}[t]
\begin{center}
\includegraphics[width=1.0\linewidth]{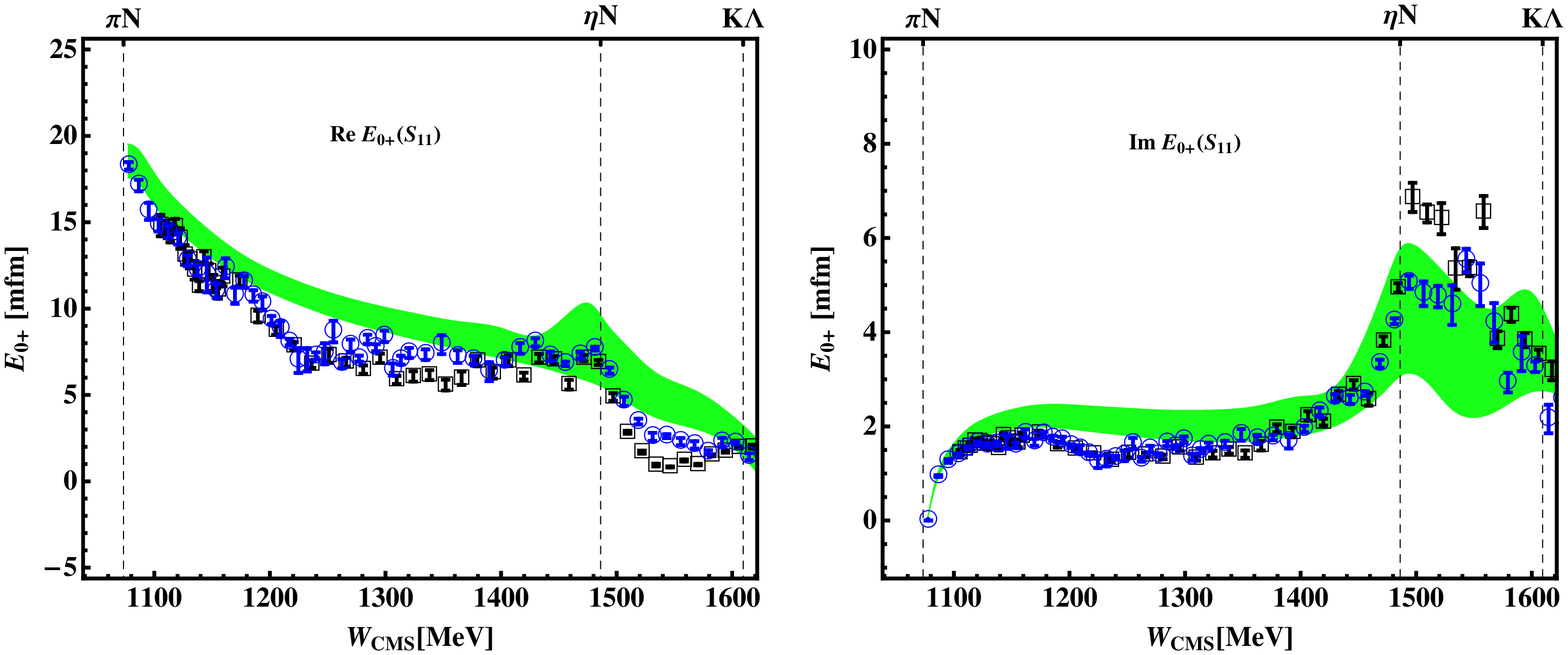}
\caption{Prediction for the multipole $E_{0+}$ for the pion photoproduction corresponding to the hadronic solution II. For comparison best fits of the MAID (circles) \cite{Drechsel:2007if} and SAID (squares) \cite{Workman:2012jf} models are represented by blue and black points with errorbars, respectively.}\label{pic:RES5new}
\includegraphics[width=1.0\linewidth]{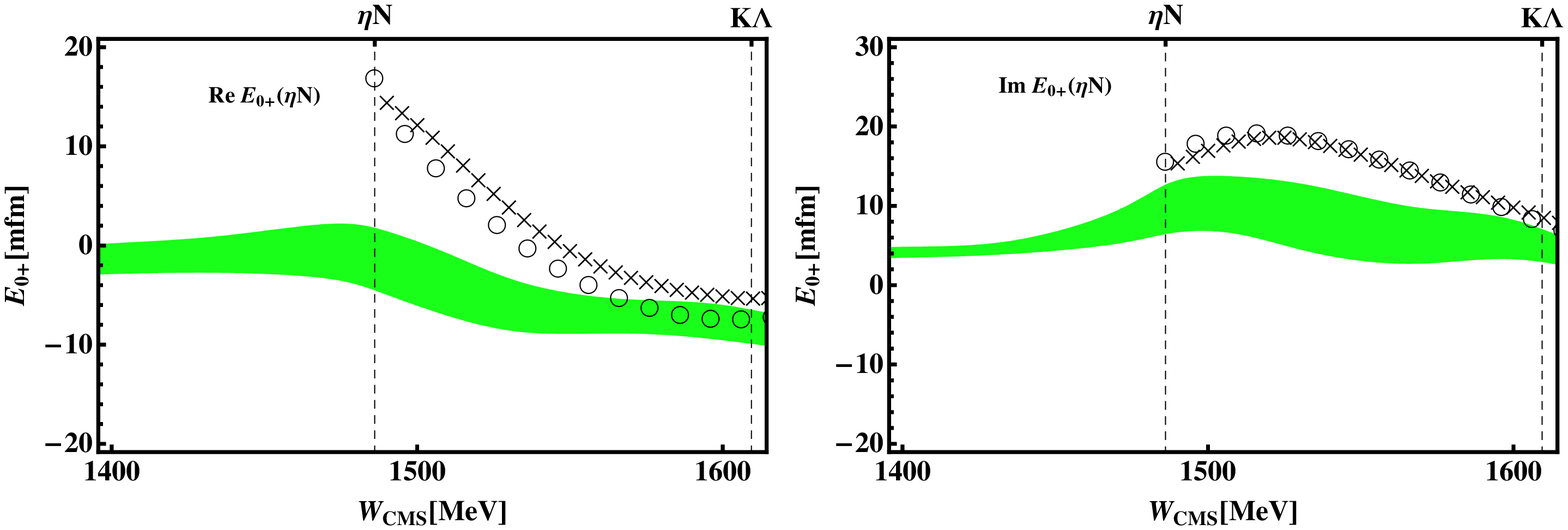}
\caption{$E_{0+}$ for the eta photoproduction as predicted starting from the hadronic solution II. For comparison we also present the outcome of the ETAMAID (circles) \cite{Chiang:2001as} and Bonn-Gatchina (squares) \cite{Anisovich:2009zy} analysis.}\label{pic:RES6new}
\end{center}
\end{figure*}

\subsection{Results}

In this section we present the prediction of the model for both sets of scattering amplitudes fixed in the hadronic sector in the last chapter. There in both strategies we have been concentrating on the description of the s-wave, thus we shall stick to the prediction of the quantities connected to this particular partial wave. Such a quantity is the electric multipole $E_{0+}$, which can be expressed in terms of the Chew, Goldberger, Low and Nambu (CGLN) amplitudes as presented in App.~\ref{app:CGLN}.

After having fixed the hadronic part of this amplitude, the photoproduction amplitude contains only 4 new parameters, namely $D$, $F$, $b_{12}$ and $b_{13}$. The first two are fixed to the commonly used values of $D=0.8$ and $F=0.5$. The 'magnetic' LECs $b_{12}$ and $b_{13}$ shall be taken from Ref.~\cite{Kubis:2000aa}, where they have been adjusted to fit the experimental data on magnetic moments. Within the uncertainty due to the choice of $m_0$ there, these LECs are given by (in units of GeV$^{-1}$)
\begin{align*}
 b_{13}=0.32\pm0.06 \quad \text{and} \quad  b_{12}=0.095\pm0.015.
\end{align*}

In order to give an impression of the uncertainty we proceed as follows. First for a fixed energy $W_{{\rm cms}}$ and for each hadronic solution which lies in the uncertainty band of the hadronic solution as presented in the last chapter we calculate the photoproduction multipoles as functions of the 4 new LECs. Then for fixed $D$ and $F$ and a large set of randomly distributed values for $b_{12}$ and $b_{13}$ within the uncertainty range on these two LECs we obtain a prediction on the photoproduction multipoles at the chosen energy. Repeating this procedure for different energy values we obtain a family of curves $E_{0+}(W_{{\rm cms}})$. The hull of all those curves is assumed to reflect the uncertainty of the model properly.

 \medskip
 
\textbf{Solution I:} 
As a central prediction of the present work we present the outcome of the multipole $E_{0+}$ for pion photoproduction in the isospin $1/2$ channel in Fig. \ref{pic:RES5}. We restrict ourselves to energies below the second nucleonic resonance as its position seems to be shifted as discussed in the last chapter. Without any fitting we observe an astonishing agreement of our prediction with the outcome of the fit from MAID2007 (updated unitary isobar model) \cite{Drechsel:2007if} and the one by the SAID group \cite{Workman:2012jf}. Of course it is clear that, if none but $\pi N$ channels are opened, Watson's theorem guarantees that the phase of $E_{0+}$ comes right, once the phase of $\pi N$ scattering has been fixed to the physical value. This theorem, however, does not fix the magnitude of the real and imaginary part of the photoproduction amplitude, neither it is clear how to apply it above the $\eta N$ threshold. 

The value of $E_{0+}$ at the threshold has been debated for a long time, see \cite{Bernard:1995dp} for a nice review on that topic. We obtain the following value
\begin{align*}
 E_{0+}^\pi (S_{11}) &= (+10.4\pm1.3) \times 10^{-3}/M_{\pi^+},
\end{align*}
which has to be compared with $E_{0+} (S_{11}) = (+12.5\pm0.3)\times
10^{-3}/M_{\pi^+}$ from experimental results \cite{Beck:1990da,Burg,
  Adamovitch} for a respective isospin combination. Seemingly our prediction
is slightly below the the experimental result. Throughout this work we have
not discussed the isospin $3/2$ channel
 and refrain from giving a numerical value of $E_{0+}$. For completeness, let
 us note that in all solutions we observe that the absolute value of the
 $E_{0+}(S_{31})$ is underestimated  significantly. We trace that discrepancy 
to the missing Born graphs which are known to be important in this channel. 
We will come back to this issue in a forthcoming work \cite{MBM_2027}.

We can go further and make a prediction on the multipole amplitude for eta photoproduction.  At the $\eta N$ threshold we extract the following value
\begin{align*}
 E_{0+}^\eta=\Big( (3.9\pm2.5) + i(10.7\pm2.7)\Big)  \times 10^{-3}/M_{\pi^+}.
\end{align*}
The energy-dependence of $E_{0+}(W_{{\rm cms}})$ is presented in Fig.~\ref{pic:RES6}, where it is compared with fits by the ETAMAID \cite{Chiang:2001as} and Bonn-Gatchina \cite{Anisovich:2009zy} groups. Seemingly there is a large qualitative agreement between our prediction and the phenomenological analysis by the ETAMAID and Bonn-Gatchina group. On a quantitative level we observe that the real part of the $E_{0+}$ is suppressed compared to the outcome of the phenomenological analysis. We wish to remind the reader that the 'magnetic' LECs are taken from a tree level calculation only. In some additional fits we have observed that the results of ETAMAID and Bonn-Gatchina group can be reproduced nicely in our approach using these LECs as free parameters. This, however, is not the original purpose of this work, namely the parameter free prediction of the photoproduction after fixing the hadronic scattering.

\medskip

\textbf{Solution II:} 
Starting from the second hadronic solution we obtain a prediction for pion photoproduction in the $S_{11}$ channel as presented in Fig. \ref{pic:RES5new}. Although all parameters of the model are fixed in the hadronic solution or taken from the literature as described above one observes a nice qualitative agreement of our prediction with the outcome of the SAID and MAID2007 analysis in a very large energy region. At the threshold we extract the following values for the lowest multipole
\begin{align*}
 E_{0+}^\pi (S_{11}) &= (+13.1\pm0.7) \times 10^{-3}/M_{\pi^+},
\end{align*}
which agrees nicely with the extraction from the experimental results in
\cite{Beck:1990da,Burg, Adamovitch}. For higher energies, i.e. around
$1200$~MeV and $1550$~MeV, we observe a discrepancy of $E_{0+}$ compared
to the fits by SAID and MAID groups. Additionally the uncertainty band appears
quite underestimated in this solution. We wish to remind the reader that the
main difference of both solutions are the three regularization scales. In the
first solution those are used as free fit parameters whereas in the second
they are fixed. The particular choice of these values is motivated as
described in the previous chapter, however, one should in principle investigate
the influence of this choice on the hadronic solution as well as on the
photoproduction amplitudes. To do so one would have to refit the hadronic
scattering for any other choice of the parameters in the solution II. Due to
an enormous amount of computational time required for each fit, we refrain
from  including that uncertainty. One should keep in mind that a more realistic uncertainty band might be larger than the one presented here.

\begin{figure}[t]
\begin{center}
\includegraphics[width=1.0\linewidth]{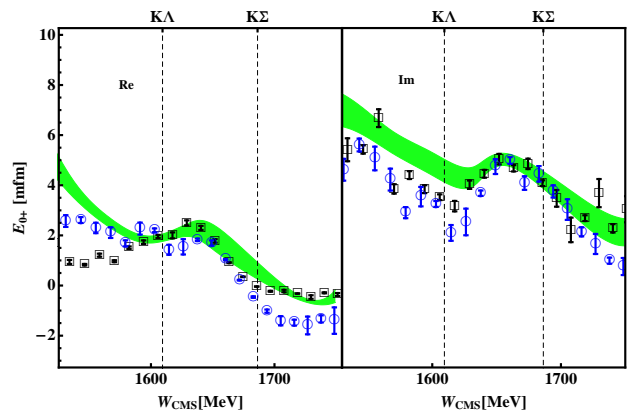}
\end{center}
\caption{A typical fit of our model to the 
SAID \cite{Workman:2012jf} (blue) and MAID \cite{Drechsel:2007if} (black)
analysis for the pion photoproduction as described in the text.}\label{pic:res5newnew}
\end{figure}

To be complete we wish to comment now on the higher energy region, i.e. above
the $K\Lambda$ threshold, where the outcome of our prediction starts to
deviate from the results of the SAID and MAID groups. In fact this observation
is identical to the one made in the analysis of the photoproduction amplitudes
in the J\"ulich model \cite{Doring:2009uc}, where no good overall fit could be
achieved for the $E_{0+}$ in the low and resonance energy region
simultaneously. Although no fit to the photoproduction data was done in in the
present work, four new parameters are entering the calculation. The axial
coupling as well as the 'magnetic' LECs $b_{12}$ and $b_{13}$ are taken from
estimations which rely on a strict perturbative calculation. Our
non-perturbative framework is on the other hand suited to extend the range of
applicability of the effective field theory. Thus it is a priori not clear
whether it is sufficient to use this new LECs in the whole energy range. To
underline this we fit our model to the SAID pion photoproduction data with
axial coupling and 'magnetic' LECs treated as free parameters. The best fit is
presented in Fig.~\ref{pic:res5newnew}, where we observe a nice agreement 
above the $K\Lambda$ threshold with the phenomenological models from SAID and MAID.

For the eta photoproduction the prediction of the second solution is presented in Fig. \ref{pic:RES6new}. At the $\eta N$ threshold we obtain the following value
\begin{align*}
 E_{0+}^\eta=\Big( (-1.2\pm2.2) + i(6.9\pm2.3)\Big)  \times 10^{-3}/M_{\pi^+},
\end{align*}
which undershoots the numerical value obtained in the previous solution for
the real part slightly and agrees for the imaginary part within the
uncertainty range. The functional behavior of $E_{0+}$ is suppressed
compared to the previous solution and even more in comparison to the ETAMAID
and Bonn-Gatchina fits. As already discussed in the previous solution we can
trace this discrepancy to the 'magnetic' LECs, which are taken from the tree
level calculation \cite{Kubis:2000aa}. These LECs do not change the functional
form of the photoproduction amplitude but seem to enhance or suppress the
structures present in the photoproduction amplitude. Those structures on the
other hand seem to reflect one-to-one the structures arising from the dynamics
of the hadronic scattering process. Thus the correct description of meson
photoproduction is necessarily to be connected to a proper description of 
the underlying hadronic scattering reactions.

\medskip

\section*{Acknowledgments}

We would like to thank M.~D\"oring, S.~Prakhov, R.~Workman and D.~Phillips for valuable discussions and comments. We also wish to acknowledge the 
great work of H. Van Pee, who takes care of the HISKP cluster. This work was supported in part
by the DFG through funds provided to SFB/TR~16 and SFB/TR~55, from the EU to
the EPOS network within HadronPhysics3 and by the Bonn-Cologne Graduate School in
Physics and Astronomy.

\begin{widetext}
\appendix

\section{Channel space structures}\label{app:coupling}

For the channel indices $\{b,j;i,a\}$ corresponding to the process
$\phi_iB_a\rightarrow\phi_jB_b$,
 the relevant coupling matrices from the leading, Eq.~\reff{eqn:LAGR0}, and
 next-to leading, Eq.~\reff{eqn:LAGR1}, order chiral Lagrangian  read
\begin{align*}
A_{WT}^{b,j;i,a}=&-\frac{1}{4F_j F_i}\langle\lambda^{b\dagger}[[\lambda^{j\dagger},\lambda^{i}],\lambda^{a}]\rangle, \\
A_{14}^{b,j;i,a}=&-\frac{2}{F_j F_i}\Big[
   b_1\Big(\langle\lambda^{b\dagger}[\lambda^{j\dagger},[\lambda^{i},\lambda^{a}]]\rangle +\langle\lambda^{b\dagger}[\lambda^{i},[\lambda^{j\dagger},\lambda^{a}]]\rangle\Big)
 + b_2\Big(\langle\lambda^{b\dagger}\{\lambda^{j\dagger},[\lambda^{i},\lambda^{a}]\}\rangle +\langle\lambda^{b\dagger}\{\lambda^{i},[\lambda^{j\dagger},\lambda^{a}]\}\rangle\Big)\\
&~~~~~~~~~~~~~~~~~~~~~~~~~~~~~~~~~~~~~~~~~~~ +b_3\Big(\langle\lambda^{b\dagger}\{\lambda^{j\dagger},\{\lambda^{i},\lambda^{a}\}\}\rangle +\langle\lambda^{b\dagger}\{\lambda^{i},\{\lambda^{j\dagger},\lambda^{a}\}\}\rangle\Big)
 + 2b_4 \langle\lambda^{b\dagger}\lambda^{a}\rangle \langle\lambda^{j\dagger}\lambda^{i}\rangle \Big],\\
A_{57}^{b,j;i,a}=&-\frac{2}{F_j F_i}\Big[
   b_5\langle\lambda^{b\dagger}[[\lambda^{j\dagger},\lambda^{i}],\lambda^{a}]\rangle
 + b_6\langle\lambda^{b\dagger}\{[\lambda^{j\dagger},\lambda^{i}],\lambda^{a}\}\rangle\\
&~~~~~~~~~~~~~~~~~~~~~~~~~~~~~~~~~~~~~~~~~~~~~~~~~~~~~~~~~~~~~~~~~~~~~~~~~~~~~~~~~~  + b_7\Big(\langle\lambda^{b\dagger}\lambda^{j\dagger}\rangle \langle\lambda^{i}\lambda^{a}\rangle-\langle\lambda^{b\dagger}\lambda^{i}\rangle \langle\lambda^{a}\lambda^{j\dagger}\rangle\Big) \Big],\\
A_{811}^{b,j;i,a}=&-\frac{1}{F_j F_i}\Big[
   b_8\Big(\langle\lambda^{b\dagger}[\lambda^{j\dagger},[\lambda^{i},\lambda^{a}]]\rangle +\langle\lambda^{b\dagger}[\lambda^{i},[\lambda^{j\dagger},\lambda^{a}]]\rangle \Big)
 + b_9\Big(\langle\lambda^{b\dagger}[\lambda^{j\dagger},\{\lambda^{i},\lambda^{a}\}]\rangle +\langle\lambda^{b\dagger}[\lambda^{i},\{\lambda^{j\dagger},\lambda^{a}\}]\rangle\Big)\\
&~~~~~~~~~~~~~~~~~~~~~~~~~~~~~~~~~~~~~~~~ + b_{10}\Big(\langle\lambda^{b\dagger}\{\lambda^{j\dagger},\{\lambda^{i},\lambda^{a}\}\}\rangle +\langle\lambda^{b\dagger}\{\lambda^{i},\{\lambda^{j\dagger},\lambda^{a}\}\}\rangle\Big)
 + 2b_{11}\langle\lambda^{b\dagger}\lambda^{a}\rangle \langle\lambda^{j\dagger}\lambda^{i}\rangle \Big],\\
A_{M}^{b,j;i,a}=&-\frac{1}{2 F_j F_i}\Big[
   2b_0\Big(\langle\lambda^{b\dagger}\lambda^{a}\rangle \langle[\lambda^{j\dagger}\lambda^{i}]\bar{\mathcal{M}}\rangle\Big)+ b_D\Big(\langle\lambda^{b\dagger}\{\{\lambda^{j\dagger},\{\bar{\mathcal{M}},\lambda^{i}\}\},\lambda^{a}\}\rangle+\langle\lambda^{b\dagger}\{\{\lambda^{i},\{\bar{\mathcal{M}},\lambda^{j\dagger}\}\},\lambda^{a}\}\rangle\Big)\\
&~~~~~~~~~~~~~~~~~~~~~~~~~~~~~~~~~~~~~~~~~~~~~~~~~~~~~~~ + b_F\Big(\langle\lambda^{b\dagger}[\{\lambda^{j\dagger},\{\bar{\mathcal{M}},\lambda^{i}\}\},\lambda^{a}]\rangle+\langle\lambda^{b\dagger}[\{\lambda^{i},\{\bar{\mathcal{M}},\lambda^{j\dagger}\}\},\lambda^{a}]\rangle\Big)	\Big],
\end{align*}
where $\lambda^{\ldots}$ denote the $3\times 3$ channel matrices (e.g. $\phi =\phi^{i}\lambda^{i}$ for the physical meson fields) and the $F_i$ are the meson decay constants  in the respective channel. Moreover, $\bar{\mathcal{M}}$ is obtained from the quark mass matrix $\mathcal{M}$ via the Gell--Mann Oakes Renner relations, and given in terms of the meson masses as follows, $\bar{\mathcal{M}}=\frac{1}{2}{\rm diag}(M_{K^+}^2 - M_{K^0}^2 + M_{\pi^0}^2, M_{K^0}^2 - M_{K^+}^2 + M_{\pi^0}^2, M_{K^+}^2 + M_{K^0}^2 - M_{\pi^0}^2)\,$.

For the channel indices $\{b,j;a\}$ corresponding to the process
$B_a\rightarrow\phi_jB_b$ the channel-space matrix is given by
\begin{align*}
A_X^{b,j;a}=-\frac{D}{\sqrt{2}F_j} \langle \lambda^{b\dagger}\{\lambda^{j\dagger}, \lambda^a\} \rangle - \frac{F}{\sqrt{2}F_j}\langle \lambda^b[\lambda^{j\dagger}, \lambda^a]\rangle.
\end{align*}

\section{Solution of the BSE with full off-shell dependence}\label{app:solBSE}

Here we present a technique for a solution of the Bethe Salpeter equation \reff{eqn:BSE} with the full off-shell dependence. This method does not rely on any approximation of the BSE which are used very often in the literature, i.e. on-shell approximation or a three-dimensional reduction of the BSE to the Lippmann-Schwinger equation. It also is applicable for any kernel with only one restriction: The interaction kernel must consist of local terms only. Thus the solution of the BSE corresponds to an infinite chain of the Feynman bubble diagrams as presented in Fig. \ref{pic:BSE}. To keep this section short we will restrict the form of the kernel to the one used in the main text of this work. Up to the next-to-leading chiral order the meson-baryon local potential is given by the following  form
\begin{align}
 V(q_2, q_1; p)           = \sum_{i=1}^6 A_i~\mathcal{D}_i(q_2, q_1; p),
\end{align}
with
\begin{align*}
\mathcal{D}(q_2, q_1; p) &= \big( \slashed{q_1},~\slashed{q_2},~(q_1\cdot q_2),~\slashed{q_2}\slashed{q_1},~\mathds{1},~\big(\slashed{q_2}(q_1\cdot p),~\slashed{q_1}(q_2\cdot p)  \big),\\
			     A &= \big( A_{WT}, A_{WT}, (A_{14}+2A_{57}),A_{57}, A_{M}, A_{811}, A_{811}      \big),  
\end{align*}
where $q_{1/2}$ and $p$ denote the four momentum of the in-/outgoing meson and the overall four momentum of the scattering process. The capital letters $A$ denote the channel space structures specified in the App.~\ref{app:coupling}, where also the $\mathds{1}$ corresponds to a unity matrix.

As already discussed the solution of the BSE corresponds to an infinite chain of Feynman diagrams which reads
\begin{align*}
 T(q_2, q_1; p)=V(q_2, q_1; p)+ i\int\frac{d^d l}{(2\pi)^d}V(\slashed{q}_2, \slashed{l}; p) \frac{\slashed p-\slashed l+m}{\big((p-l)^2-m^{2}\big)\big(l^2-M^2\big)}V(\slashed{l}, \slashed{q}_1; p)+...~,
\end{align*}
where we again have suppressed the channel indices keeping in mind that $T$
and $V$ are matrices in channel space. 
From the last equation one easily sees that the iterative use of the interaction potential introduces new Dirac-momentum structures additionally to those of $\mathcal{D}$. The number of those is limited ensuring that one can separate the channel space structures from the Dirac-momentum structures of the scattering matrix as follows
\begin{align*}
 T(q_2, q_1; p)= \sum_{i=1}^{20} {\rm ~T}_i(s)~\aleph_i(q_2, q_1; p)~,
\end{align*}
with
\begin{align*}
\aleph(q_2, q_1; p):=&\big(\slashed{q_1},~\slashed{p}\slashed{q_1},~\slashed{q_2}\slashed{p}\slashed{q_1},~\slashed{q_2}\slashed{q_1},~\slashed{p}\slashed{q_1}(q_2\cdot p),~\slashed{q_1}(q_2\cdot{}p),~\slashed{q_2}(q_1\cdot p),~\slashed{q_2}\slashed{q_1},~(q_1\cdot p)(q_2\cdot p),  \\ 
&~\slashed{p}(q_1\cdot~p)(q_2\cdot~p),~(q_1\cdot p){}, \slashed{p}(q_1\cdot p),~(q_2\cdot q_1), \slashed{p}(q_2\cdot q_1),~\slashed{q_2}\slashed{p},~\slashed{q_2}{},~\slashed{p}(q_2\cdot p),~(q_2\cdot p),~\mathds{1},~\slashed{p}\big).
\end{align*}
Please note that different to the decomposition of the potential $V$ the
coefficients ${\rm T}_i(s)$ still depend on the center-of-mass energy squared,
i.e. $s$. The Dirac-momentum space spanned by the vectors $\mathcal{D}$ is a
subspace of those spanned by the vectors $\aleph$. 
Consequently, the interaction potential $V$ can also be expressed in terms of these vectors. More importantly 
\begin{align*}
 \forall a\in\aleph(q_2, l; p), b\in\aleph(l, q_1; p): \qquad \int\frac{d^d l}{(2\pi)^d}  \frac{a(\slashed p-\slashed l+m)b}{\big((p-l)^2-m^{2}\big)\big(l^2-M^2\big)} \in \aleph(q_2, q_1; p).
\end{align*}
This ensures that Eq.~\reff{eqn:BSE} can be rewritten in a linear equation of the form
\begin{align}\label{solution}
 {\rm X}^i_{~j}(s){\rm T}^j(s)=V^i
\text{ for } i,j=1,..,20 \text{ and } {\rm X }\in Mat^{\aleph(q_2, q_1; p)\times\aleph(q_2, q_1; p)}.
\end{align}
We refrain from presenting the matrix $\rm X$ as well as the solution of
Eq.~\reff{solution} explicitly for 
reasons of brevity but it should be mentioned that every element is a function
of the baryon/meson masses, the c.m. energy squared as well as of the scalar
loop integrals $I_M$ and $I_{MB}$, which are collected in the App.~\ref{app:loop}. Moreover every element of $\rm X$ is a matrix in the channel space which depends on the LECs as presented in App. \ref{app:coupling}. The latter implies a non-commuting character of matrix elements ${\rm X}^i_{~j}$, which one should keep in mind while solving Eq.~\reff{solution}.

\section{Loop integrals}\label{app:loop}

Here we collect all loop integrals required for the calculation of the scattering as well as the photoproduction amplitudes. Note that for reasons given in the main part all purely baryonic integrals are set to zero from the beginning. Utilizing dimensional regularization in the $\overline{MS}$ scheme the one-meson integral is given by
\begin{align*}
 I_M(M):=&\mathop{\int}_{\overline{MS}} \frac{d^dl}{(2\pi)^d}\frac{i}{l^2-M^2+i\epsilon}\stackrel{d=4}{=}\frac{1}{16 \pi^2}\Big[2M^2\log\big(\frac{M}{\mu}\big)\Big],
\end{align*}
where $\mu$ is the regularization scale and $M$ denotes the meson mass. We use in the following the common abbreviation $\lambda(a,b,c)=a^2+b^2+c^2-2ab-2ac-2bc$, such that the meson-baryon (of masses $M$ and $m$, respectively) integral reads
\begin{align*}
 I_{MB}(s,M,m):=&\mathop{\int}_{\overline{MS}}\frac{d^dl}{(2\pi)^d}\frac{1}{l^2-M^2+i\epsilon}\frac{i}{(l-p)^2-m^2+i\epsilon}\\
              \stackrel {d=4}{=}&\frac{1}{16 \pi^2}\Big[-1 +  2\log\big(\frac{m}{\mu}\big)+ \frac{M^2- m^2 + s}{s}\log\big(\frac{M}{m}\big)-2 \frac{\sqrt{\lambda(s,M,m)}}{s}\mathrm{arctanh}\Big(\frac{\sqrt{\lambda(s,M,m)}}{(m + M)^2 - s}\Big)\Big].
\end{align*}
The photoproduction amplitude involves further loop integrals. The triangle graph of class ``D'' in the Fig. \ref{pic:photo} gives rise to a meson-meson-baryon as well as via the Passarino-Veltman reduction to a meson-meson loop integral at $s=k^2$, which read
\begin{align*}
 I_{MM}(k^2,M):=&\mathop{\int}_{\overline{MS}} \frac{d^dl}{(2\pi)^d}\frac{1}{l^2-M^2+i\epsilon}\frac{i}{(l-k)^2-M^2+i\epsilon}\\
              \stackrel{d=4}{=}&\frac{1}{16 \pi^2}\Big[-2 +  2\log\big(\frac{M}{\mu}\big) -2 \frac{\sqrt{\lambda(s,M,M)}}{s}\mathrm{arctanh}\Big(\frac{\sqrt{\lambda(s,M,M)}}{4M^2 - s}\Big)\Big],\\
 I_{MMB}(s,k^2,M,m):=&\mathop{\int}_{\overline{MS}} \frac{d^dl}{(2\pi)^d} \frac{1}{l^2-M^2+i\epsilon}  \frac{1}{(l-k)^2-M^2+i\epsilon}  \frac{i}{(l-k-p_1)^2-m^2+i\epsilon}~,
\end{align*}
where $p_1=p-q'$ is the four-momentum of the incoming proton. We wish to emphasize that the photon coupled to a meson propagator does not induce a transition of this meson. Differently, coupled to a baryon propagator it can induce the $\Sigma^0 \leftrightarrow \Lambda$ transition. Thus a meson-baryon-baryon loop integral required for the calculation of Feynman diagrams of class ``E'' reads in general
\begin{align*}
 I_{MBB}(s,k^2,M,m_1,m_2):=&\mathop{\int}_{\overline{MS}} \frac{d^dl}{(2\pi)^d} \frac{1}{l^2-M^2+i\epsilon}  \frac{i}{(l-p_1)^2-m_1^2+i\epsilon}  \frac{i}{(l-k-p_1)^2-m_2^2+i\epsilon}~.\\             
\end{align*}
Both last integrals can not be written in terms of elementary functions. We solve both integrals utilizing Cutkosky rules to calculate the imaginary part of the meson-baryon-baryon integral. A non-subtracted dispersion relation then gives the real part of the loop integral as follows,
\begin{align*}
 \Im(I_{MBB}(s,k^2,M,m_1,m_2))&=\frac{1}{32\pi k_{{\rm cms}}\sqrt{s}}\log\Big( \frac{m_1^2-m_2^2+k^2-2k_0q_0-2q_{{\rm cms}}k_{{\rm cms}}}{m_1^2-m_2^2+k^2-2k_0q_0+2q_{{\rm cms}}k_{{\rm cms}}}   \Big),\\ 
 \Re(I_{MBB}(s,k^2,M,m_1,m_2))&=\frac{1}{\pi}\int_{(m_2+M)^2}^\infty ds'\frac{\Im(I_{MBB}(s',k^2,M,m_1,m_2))}{s'-s}  ,            
\end{align*}
where $k_{{\rm cms}}=\sqrt{\lambda(s,p_1^2,k^2)}/(2\sqrt{s})$, $q_{{\rm cms}}=\sqrt{\lambda(s,m_1^2,M^2)}/(2\sqrt{s})$ and $k_0=\sqrt{k_{{\rm cms}}^2+k^2}$, $q_0=\sqrt{q_{{\rm cms}}^2+M^2}$. The same holds for the $I_{MMB}$, where in the last formulas one has to replace: $m_1\rightarrow M$ and $m_2\rightarrow m$.

\section{Partial wave analysis of $\pi N \rightarrow \eta N$ scattering}\label{app:pwa}
The pion-induced eta production off the neutron is dominated by the
contribution of the nearby nucleon resonances, i.e. the $S_{11}(1535)$ and
$D_{13}(1520)$. From the previous study of elastic $\pi N$ scattering we
already know that the first one is described perfectly as a dynamically
generated resonance within our approach, whereas the d-wave resonance is
not. Thus we wish to clarify, whether an ansatz for the scattering amplitudes,
which contains s- and p-waves only, is capable to generate a
$\cos^{2}(\theta)$-like behavior of the differential cross sections
$d\sigma/d\Omega(s,\cos(\theta))$ ($\theta$ here denotes the scattering angle
in the c.m. frame). As a matter of fact this generation of a
$\cos^{2}(\theta)$-like structure through the iteration of p-waves does not
seem to be appreciated
in several experimental works, see e.g. \cite{Prakhov:2005qb}. There, the
presence of 
a $\cos^{2}(\theta)$-behavior in the shape of differential cross section is assumed to be a direct indication for a d-wave dominance.

Let us start from the most general form of the T-matrix, which is invariant under Lorentz as well as parity transformations. For the scattering of a meson-baryon system from initial state $(i)$ with the meson momentum $(q)$, and baryon momentum and spin $(p, s)$ to the final state $(f)$ with meson momentum $(q')$, and baryon momentum and spin $(p',s')$ it reads with the usual conventions used by H\"ohler \cite{Hoehler}
\begin{align*}
 M_{fi}=\frac{1}{8\pi\sqrt{s}}\bar u_f(p',s')\{A_{fi}(s,t)+\frac{1}{2}(\slashed{q}+\slashed{q}')B_{fi}(s,t)\} u_i(p,s),
\end{align*}
where $s=P^2:=(p+q)^2=(p'+q')^2$ and $t=(q-q')^2=(p-p')^2$ are the Mandelstam variables. The amplitudes $A$ and $B$ can be recombined to the scattering amplitude on the mass shell $T_{ON}(\slashed{q}',\slashed{q}; P)$ as follows
\begin{align*}
 T_{ON}(\slashed{q}',\slashed{q}; P)=T_{ON}^0(s,\textbf{z})+\slashed{P}T_{ON}^1(s,\textbf{z}) = A(s,t)+\frac{1}{2}(\slashed{q}+\slashed{q}')B(s,t).
 \end{align*}
Here, $z=\cos(\theta)$ is the standard representation of the scattering angle.
In fact, $z$  is related to the Mandelstam $t$ via
\begin{align*}
t=M_f^2+M_i^2-2\sqrt{q_i^2+M_i^2}\sqrt{q_f^2+M_f^2}+2q_iq_f\, {z}~,
\end{align*}
where $q_{i/f}$ is the modulus of the center of mass momentum of the in- and outgoing system respectively. Suppressing the kinematic variables for the moment the differential cross section in the center of mass system reads
\begin{align*}
\Big(\frac{d\sigma(i\rightarrow f)}{d\Omega}\Big)=\frac{1}{64\pi^2}\frac{q_f}{q_i}\frac{1}{2}\sum_{s,s'}\Big|\bar u_f(p',s')\{T_{ON}^{0;fi}+\slashed{P}T_{ON}^{1;fi}\} u_i(p,s)\Big|^2.
\end{align*}
For Dirac spinors normalized such that $\bar u_f(p)u_i(p)=2m\delta_{fi}$, and suppressing for brevity the channel indices, the spin sum can be calculated in terms of $T_{ON}^0$ and $T_{ON}^1$ as follows
\begin{align*}
\frac{1}{2}\sum_{s,s'}\Big|\bar u_f(p',s')\{T_{ON}^0+\slashed{P}T_{ON}^1\} u_i(p,s)\Big|^2=c_{00}|T_{ON}^0|^2+2c_{01}\Re(T_{ON}^{1*}T_{ON}^{0}) + c_{11}|T_{ON}^1|^2,
\end{align*}
where
\begin{align*}
 c_{00}&=\frac{1}{2s}\Big( (s+m_i^2-M_i^2)(s+m_f^2-M_f^2) + 4s(m_im_f-zq_fq_i)    \Big),\\
 c_{01}&=m_i(s+m_f^2-M_f^2)+m_f(s+m_i^2-M_i^2),\\
 c_{11}&=\frac{1}{2}\Big( (s+m_i^2-M_i^2)(s+m_f^2-M_f^2) + 4s(m_im_f+zq_fq_i)    \Big).
\end{align*}

The above formulae specify all required kinematics and spin structure. The dynamical input is incorporated within the scattering amplitudes, $A$ and $B$. In the main body of this work these are taken to be solutions of the BSE. In view of the above question, we wish to make an ansatz for the scattering amplitudes. First of all, the standard amplitudes $A$ and $B$ can be expanded in Legendre polynomials $P_l(z)$ as follows
\begin{align*}
\frac{A(s,t)}{4\pi}&=\frac{\sqrt{W_{{\rm cms}}+m_i}}{\sqrt{E_{{\rm cms};i}+m_i}}f_1(s,t)\frac{\sqrt{W_{{\rm cms}}+m_f}}{\sqrt{E_{{\rm cms};f}+m_f}} - \frac{\sqrt{W_{{\rm cms}}-m_i}}{\sqrt{E_{{\rm cms};i}-m_i}}f_2(s,t)\frac{\sqrt{W_{{\rm cms}}-m_f}}{\sqrt{E_{{\rm cms};f}-m_f}},\\
\frac{B(s,t)}{4\pi}&=\frac{1}{\sqrt{E_{{\rm cms};i}+m_i}}f_1(s,t)\frac{1}{\sqrt{E_{{\rm cms};f}+m_f}}+\frac{1}{\sqrt{E_{{\rm cms};i}-m_i}}f_2(s,t)\frac{1}{\sqrt{E_{{\rm cms};f}-m_f}},
\end{align*}
where $W_{{\rm cms}}=\sqrt{s}$ and $E_{{\rm cms};i/f}=\sqrt{q_{i/f}^2+m_{i/f}^2}$. After a variable transformation the amplitudes $f_{1,2}$ are related to the commonly used partial wave amplitudes $f_{l\pm}(s)$ as follows \cite{Hoehler,Chew:1957zz},
\begin{align*}
f_1(s,z)&=\sum_{l=1}^\infty(f_{(l-1)+}(s)-f_{(l+1)-}(s))P_l'(z),\\
f_2(s,z)&=\sum_{l=1}^\infty(f_{l-}(s)-f_{l+}(s))P_l'(z).
\end{align*}
For the purpose of this section we do not consider  additional constraints for
the partial wave amplitudes, e.g. 
due to analyticity or unitarity. Thus both real and imaginary part of those are used as free parameters, which will be adjusted to reproduce the data on differential cross sections for the process $\pi^-N\rightarrow \eta N$, measured by Prakhov et al., see Ref.~\cite{Prakhov:2005qb}. For the truncation of the partial wave expansion we assume three different scenarios:

\begin{figure}
 \includegraphics[width=0.99\linewidth]{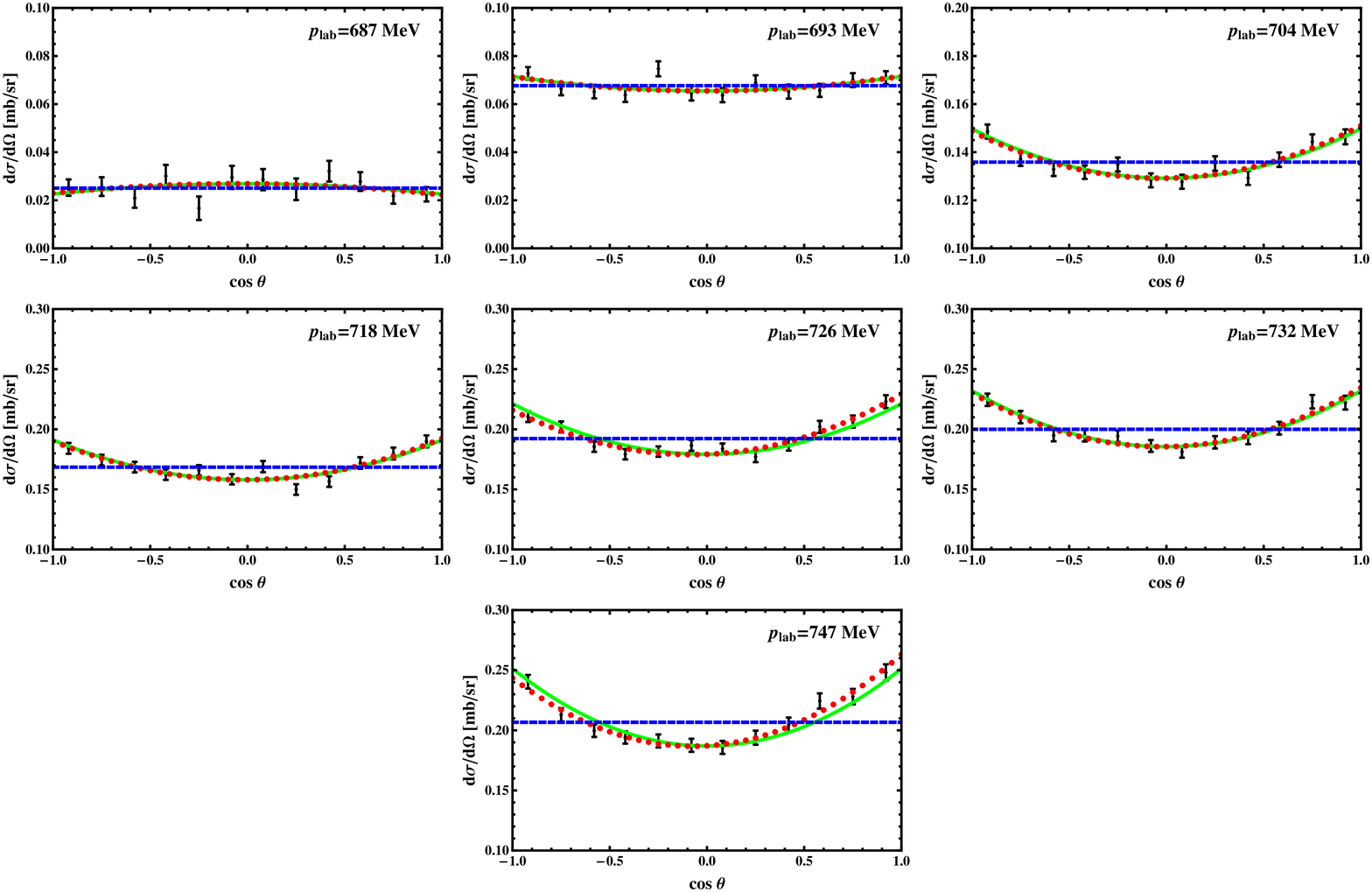}
 \caption{Best fits of three assumed scenarios to the data of the differential
   cross sections from Ref.~\cite{Prakhov:2005qb} 
  for different pion momenta $p_{\rm lab}$. The dashed (blue), red (dotted)
  and green (full) line correspond  to  the first, second and third scenario, respectively, as described in the text.} \label{picapp:pwa}
\end{figure}

\begin{enumerate}
\item The scattering amplitude contains only the s-wave dynamically, meant in the sense of the above discussion. Free parameters are $\{\Re(f_{0+}),\Im(f_{0+})\}$.
\item Both, s- and p-waves are included dynamically. This is the case for the solution of the BSE with contact terms from NLO chiral Lagrangian as performed in our approach. Without restricting the parity of the p-waves we end up with the following free parameters  for this scenario: $\{\Re(f_{0+}),\Im(f_{0+}),\Re(f_{1-}),\Im(f_{1-}),\Re(f_{1+}),\Im(f_{1+})\}$.
\item The scattering amplitude is determined by s- and d-wave (for instance $D_{13}$), whereas the p-wave is negligible. This is the case for the process $\pi N\rightarrow \eta N$ from the phenomenological point of view. For phenomenological reasons we assume only the $D_{13}$ wave to be non-negligible. Thus the free parameters are $\{\Re(f_{0+}),\Im(f_{0+}),\Re(f_{2-}),\Im(f_{2-})\}$.
\end{enumerate}

For each incident $\pi^-$ momenta separately and for each scenario we obtain best fits as presented in Fig.~\ref{picapp:pwa}. As expected the first scenario is only capable to fit the data at lowest beam momenta. The s-wave is dominant at low energies, however at higher energies it lacks the angular dependence and thus fails to describe the data properly. Going to higher beam momenta both, the second and third scenario describe the data equally well. It turns out that the presence of p-waves of both parities is required to reproduce the $z^2$-behavior. Thus albeit our approach based on the unitarization of the NLO chiral potential does not produce d-waves in the sense of the above discussion, it is capable to reproduce the data on differential cross section well enough.

\section{One-photon vertices}\label{app:ver}

For all vertices the in(out)going meson and baryon are denoted by the channel index $i(j)$ and $a(b)$ respectively, whereas the charge of the corresponding particle is denoted by $Q$. All required photon induced vertices $W$ from the leading order chiral Lagrangian Eq.~\reff{eqn:LAGR0} read
\begin{align*}
 W^{\mu~~~b;a}_{\gamma B\rightarrow B}        &=   ieQ_a	\cdot \gamma^\mu~, \qquad&                  W^{\mu~~~j;i}_{\gamma \phi\rightarrow \phi}  &=   ieQ_i	\cdot (2q_2^\mu - k^\mu)~,\\
 W^{\mu~~~b,j;a}_{\gamma B\rightarrow \phi B} &=   ieQ_jA_x^{b,j;a}	\cdot\gamma^\mu\gamma^5~, \qquad&   W^{\mu~~~b,j;i,a}_{\gamma \phi B\rightarrow \phi B} &= -ie(Q_i+Q_j)A_{WT}^{b,j;i,a}	\cdot\gamma ^{\mu },
\end{align*}
where $q_2$ denotes the four-momentum of the produced (outgoing) meson. The vertices from the second order chiral Lagrangian \reff{eqn:LAGR1} posses more involved channel structures, which for instance can be traced back elegantly to the already defined channel matrices in \ref{app:coupling}. The interaction vertex for the process $\gamma(k)\phi_i(q_1)B_a(p_1)\rightarrow\phi_j(q_2)B_b(p_2)$ reads 
\begin{align*}
 W^{\mu~~~b,j;i,a}_{\gamma \phi B\rightarrow \phi B} &= ~~~  iA_{14}\cdot \Big(Q_jq_1^\mu +Q_iq_2^\mu\Big)\\
						     &  ~~~+ iA_{57} \cdot \Big( Q_j[\gamma^\mu,\slashed{q_1}]+ Q_i[\gamma^\mu,\slashed{q_2}] \Big) \\
						     &  ~~~+ iA_{811} \cdot \Big(Q_j \gamma^\mu (q_1,p_1+q_1) + Q_j\slashed{q_1}(q_1^\mu+p_1^\mu) + Q_i \gamma^\mu (q_2,q_2+p_2) + Q_i\slashed{q_2}(q_2^\mu+p_2^\mu) + \slashed{q_1}q_1^\mu+\slashed{q_2}q_2^\mu\Big)\\
						     &  ~~~-\frac{ie}{2 F_iF_j} \Big(
							     b_{12} \langle \lambda^{b\dagger}[Q_i[\lambda^i,\lambda^{j\dagger}] - Q_j[\lambda^{j\dagger},\lambda^i],\lambda^a]\rangle
						           + b_{13} \langle \lambda^{b\dagger}\{Q_i[\lambda^i,\lambda^{j\dagger}] - Q_j[\lambda^{j\dagger},\lambda^i],\lambda^a\}\rangle \Big) \cdot (\slashed{k}\gamma^\mu-\gamma^\mu\slashed{k}).
\end{align*}
The latter expression originates from the electromagnetic field-strength tensor $f_+^{\mu\nu}$. Furthermore the same term gives rise to an additional coupling of the photon to a baryon, which does not vanish for electrically neutral baryons. It also induces a baryon transition $\Sigma^0\leftrightarrow\Lambda$, the corresponding vertex reads
\begin{align*}
 W^{\mu~~~b;a}_{\gamma B\rightarrow B} &= 2 ie \Big( b_{12} \langle \lambda^{b\dagger}[Q,\lambda^a] \rangle + b_{13} \langle \lambda^{b\dagger}\{Q,\lambda^a\}\rangle \Big)   \cdot (\slashed{k}\gamma^\mu-\gamma^\mu\slashed{k}),
\end{align*}
where $Q={\rm diag}(2/3,-1/3,-1/3)$ is the charge matrix and $e$ is the charge of an electron.

\section{Multipoles}\label{app:CGLN}

In this section we wish to specify the major technical steps on the way from the photoproduction amplitude as calculated utilizing usual Feynman rules to the multipole amplitudes as well as to the cross sections. In large parts of this section we use the conventions of \cite{Berends:1967vi} and start from the most general Lorentz covariant transition matrix element for the process of meson $(\phi_f)$ production of the baryon ($B_i$) via an incoming photon $(\gamma (k))$, i.e. $\gamma(k)B_i(p-k)\rightarrow B_f(p-q)\phi_f(q)$. It reads
\begin{align}
 T_{fi}=i\epsilon_\mu \, \bar u_f (\sum_{k=1}^8\mathcal{B}_k\mathcal{N}_k^\mu)u_i,
\end{align}
where $\epsilon_\mu$ is the photon polarization vector. The initial and final
Dirac spinors $u_i$ and $u_f$ 
are normalized like $\bar u u=2m$, with $m$ the mass of the corresponding baryon. The coefficients $\mathcal{B}_i$ are functions of the coefficients of the hadronic scattering amplitude $\{{\rm T}_i; i=1,...,20\}$ as defined in the App.~\ref{app:solBSE}, loop integrals from App.~\ref{app:loop} as well as vertices from App.~\ref{app:coupling}. Since both baryons are on-shell there are only 8 different structures, i.e. $\mathcal{N}^\mu_i~\in~\{\gamma_5\gamma^\mu\slashed k,2\gamma_5 P^\mu,2\gamma_5q^\mu,2\gamma_5k^\mu,\gamma_5\gamma^\mu,\gamma_5\slashed k P^\mu ,\gamma_5\slashed k k^\mu,\gamma_5\slashed k q^\mu \}$ with $P=\frac{1}{2}(2p-q-k)$.

Fixing the axis of quantization to the z-axis, one is able to reduce the Dirac spinors to the two-component spinors $\chi$ as follows
\begin{align}
 T_{fi}=8\pi\sqrt{s}~\chi_f^\dagger \sum_{k=1}^8\mathcal{F}_k\mathcal{G}_k\chi_i~.
\end{align}
This gives rise to the so-called Chew, Goldberger, Low and Nambu (CGLN) \cite{Chew:1957zz}
amplitudes $\mathcal{F}_i$, which are defined in the basis given by 
$\mathcal{G}_k~\in~\{i(\stackrel{\rightarrow}{\sigma} \cdot \stackrel{\rightarrow}{\epsilon}),
(\stackrel{\rightarrow}{\sigma} \cdot\hat q)(\stackrel{\rightarrow}{\sigma}\cdot[\hat k\times\stackrel{\rightarrow}{\epsilon}]),i(\stackrel{\rightarrow}{\sigma} \cdot\hat k)(\hat q\cdot \stackrel{\rightarrow}{\epsilon}),
i(\stackrel{\rightarrow}{\sigma} \cdot\hat q)(\hat q\cdot \stackrel{\rightarrow}{\epsilon}),i(\stackrel{\rightarrow}{\sigma} \cdot \hat k )(\hat k \cdot \stackrel{\rightarrow}{\epsilon}),
i(\stackrel{\rightarrow}{\sigma} \cdot\hat q) (\hat k\cdot \stackrel{\rightarrow}{\epsilon}),i(\stackrel{\rightarrow}{\sigma} \cdot \hat q)(\hat k\cdot \stackrel{\rightarrow}{\epsilon}),
i(\stackrel{\rightarrow}{\sigma} \cdot\hat
q)\epsilon_0,i(\stackrel{\rightarrow}{\sigma} \cdot\hat k)\epsilon_0 \}$. 
Here, an arrow denotes a three-dimensional vector and a hat a normalized
three-vector. Due to current conservation, 
two of the eight CGLN amplitudes can be eliminated via
\begin{align*}
 \mathcal{F}_1+ (\hat k \cdot \hat q)\mathcal{F}_3+\mathcal{F}_5-\frac{k_0}{|k|}\mathcal{F}_8 = 0 \quad \text{and} \quad (\hat k \cdot \hat q)\mathcal{F}_4 + \mathcal{F}_6 -\frac{k_0}{|k|}\mathcal{F}_7 =0,
\end{align*}
which serves as a good check of our calculation. To further extent two of the remaining six amplitudes are accompanied by scalar components of $\epsilon$ only and thus have no influence on photoproduction amplitudes, i.e. process including real photons. Finally the lowest electric multipole $E_{0+}$ can be calculated as follows
\begin{align}
 E_{0+}=\int_{-1}^1 dz \Big( \frac{1}{2} P_0\mathcal{F}_1 - \frac{1}{2} P_1\mathcal{F}_2+ \frac{1}{6} (P_0-P_2)\mathcal{F}_2 \Big),
\end{align}
where $P_l$ denote the Legendre polynomials. The latter as well as the CGLN amplitudes are functions of the cosine of the scattering angle in the c.m. system, $z$. The unpolarized differential cross section for meson photoproduction is given by
\begin{align}
 \frac{d\sigma}{d\Omega}=\frac{|q|}{|k|}\Big(&~ |\mathcal{F}_1|^2 + |\mathcal{F}_2|^2 + \frac{1}{2}|\mathcal{F}_3|^2 + \frac{1}{2}|\mathcal{F}_4|^2 + \Re(\mathcal{F}_1\mathcal{F}_4^*)+\Re(\mathcal{F}_2\mathcal{F}_3^*)+ z\Re(\mathcal{F}_3\mathcal{F}_4^*-2\mathcal{F}_1\mathcal{F}_2^*) \nn\\
                         &-z^2(\frac{1}{2}|\mathcal{F}_3|^2+\frac{1}{2}|\mathcal{F}_4|^2+\Re(\mathcal{F}_1\mathcal{F}_4^*+\mathcal{F}_2\mathcal{F}_3^*))-z^3\Re(\mathcal{F}_3\mathcal{F}_4^*) \Big).
\end{align}

\smallskip

\end{widetext}


\end{document}